\documentclass[
reprint,
superscriptaddress,
amsmath,amssymb,
aps,
prb,
prstab,
floatfix,
longbibliography
]{revtex4-2}

\usepackage[caption=false]{subfig}
\usepackage{multirow}
\usepackage{graphicx}
\usepackage{dcolumn}
\usepackage{bm}
\usepackage{threeparttable}
\usepackage{color}
\usepackage[dvipsnames]{xcolor}
\usepackage{chemformula}
\usepackage{xspace}
\usepackage{orcidlink}
\usepackage{hyperref}
\usepackage{booktabs}
\usepackage{mathtools}
\usepackage{array,multirow,varwidth}
\usepackage{amsmath}
\usepackage[english]{babel}
\DeclareMathOperator\arctanh{arctanh}

\definecolor{newtext}{RGB}{0, 0, 0}

\hypersetup{colorlinks=true,linkcolor=blue,filecolor=blue,urlcolor=blue,citecolor=blue,pdftitle = {Spin and lattice dynamics at the spin-reorientation transitions in the rare-earth orthoferrite Sm$_{0.55}$Tb$_{0.45}$FeO$_{3}$}, pdfauthor = {R.~M.~Dubrovin}}

\newcommand{\TFO}{$\mathrm{TbFeO}_{3}$}
\newcommand{\SFO}{$\mathrm{SmFeO}_{3}$}
\newcommand{\STFO}{$\mathrm{Sm}_{0.55}\mathrm{Tb}_{0.45}\mathrm{FeO}_{3}$}
\newcommand{\RFO}{$\mathrm{RFeO}_{3}$}

\definecolor{RVM}{RGB}{228, 26, 28}

\bibpunct{[}{]}{,}{n}{}{}

\begin{document}

\preprint{APS/123-QED}

\title{Spin and lattice dynamics at the spin-reorientation transitions\\ in the rare-earth orthoferrite Sm$_{0.55}$Tb$_{0.45}$FeO$_{3}$}

\author{R.~M.~Dubrovin\,\orcidlink{0000-0002-7235-7805}}
\email{dubrovin@mail.ioffe.ru}
\affiliation{Ioffe Institute, Russian Academy of Sciences, 194021 St.\,Petersburg, Russia}

\author{A.~I.~Brulev\,\orcidlink{0009-0004-5339-8486}}
\affiliation{Ioffe Institute, Russian Academy of Sciences, 194021 St.\,Petersburg, Russia}
\affiliation{University of Nizhny Novgorod, 603022 Nizhny Novgorod, Russia}

\author{N.~R.~Vovk\,\orcidlink{0000-0001-8697-4452}}
\affiliation{Department of Physics, Lancaster University, Bailrigg, Lancaster LA1 4YW, United Kingdom}

\author{I.~A.~Eliseyev\,\orcidlink{0000-0001-9980-6191}}
\affiliation{Ioffe Institute, Russian Academy of Sciences, 194021 St.\,Petersburg, Russia}

\author{N.~N.~Novikova\,\orcidlink{0000-0003-2428-6114}}
\affiliation{Institute of Spectroscopy, Russian Academy of Sciences, 108840 Moscow, Troitsk, Russia}

\author{V.~A.~Chernyshev\,\orcidlink{0000-0002-3106-3069}}
\affiliation{Department of Basic and Applied Physics, Ural Federal University, 620002 Yekaterinburg, Russia}

\author{A.~N.~Smirnov\,\orcidlink{0000-0001-9709-5138}}
\affiliation{Ioffe Institute, Russian Academy of Sciences, 194021 St.\,Petersburg, Russia}

\author{V.~{Yu}.~Davydov\,\orcidlink{0000-0002-5255-9530}}
\affiliation{Ioffe Institute, Russian Academy of Sciences, 194021 St.\,Petersburg, Russia}

\author{Anhua~Wu\,\orcidlink{0000-0002-0979-8085}}
\affiliation{Shanghai Institute of Ceramics, Chinese Academy of Sciences, 201899 Shanghai, China}

\author{Liangbi~Su\,\orcidlink{0000-0002-5792-1797}}
\affiliation{Shanghai Institute of Ceramics, Chinese Academy of Sciences, 201899 Shanghai, China}

\author{R.~V.~Mikhaylovskiy\,\orcidlink{0000-0003-3780-0872}}
\affiliation{Department of Physics, Lancaster University, Bailrigg, Lancaster LA1 4YW, United Kingdom}

\author{A.~M.~Kalashnikova\,\orcidlink{0000-0001-5635-6186}}
\affiliation{Ioffe Institute, Russian Academy of Sciences, 194021 St.\,Petersburg, Russia}

\author{R.~V.~Pisarev\,\orcidlink{0000-0002-2008-9335}}
\affiliation{Ioffe Institute, Russian Academy of Sciences, 194021 St.\,Petersburg, Russia}

\date{\today}

\begin{abstract}

Linear and non-linear couplings of magnetic and lattice excitations are at the heart of many fascinating magnetophononic phenomena observed in rare-earth orthoferrites, the distinctive feature of which is the tendency to spin-reorientation transitions.
Here we report the results of the experimental study of the spin and lattice dynamics in the Brillouin zone center of the rare-earth orthoferrite \STFO{} by using polarized infrared reflectivity and Raman scattering spectroscopic techniques.
The obtained results were supported by the first-principles calculations, which allowed us to reliably identify the parameters of most infrared- and Raman-active phonons.
We reveal the spin-reorientation transitions $\Gamma_{4}(G_{a}F_{c}) \xleftrightarrow{T_{1}} \Gamma_{24}(G_{ac}F_{ac}) \xleftrightarrow{T_{2}} \Gamma_{2}(G_{c}F_{a})$ at $T_{1} \simeq 220$\,K and $T_{2} \simeq 130$\,K and carefully studied the following evolution of Raman scattering on magnetic excitations at these transitions.
Notably, the intermediate magnetic structure $\Gamma_{24}$ displays an exceptionally broad temperature range $\Delta{T} = T_{1} - T_{2} \simeq 90$\,K in mixed \STFO{} compared to pure rare-earth orthoferrites.
We attribute this broadening of the intermediate phase to the modification of the magnetocrystalline anisotropy as a result of the inhomogeneous magnetic structure caused by the random distribution of rare-earth $\mathrm{Sm}^{3+}$ and $\mathrm{Tb}^{3+}$ ions.
We found neither change in the parameters of Raman-active $B_{1g}$ phonons nor the appearance of new phonons induced by spin-reorientation transitions, which have been reported in $\mathrm{SmFeO}_{3}$.
We assume that our results provide a solid basis for more deeper understanding 
of magnetophononic phenomena in rare-earth orthoferrites.

\end{abstract}

\maketitle

\section{Introduction}

Orthoferrites $\mathrm{RFeO}_{3}$, where $\mathrm{R}$ stands for a rare-earth cation, have been known for over 60 years and are at the forefront of modern magnetism research throughout that time~\cite{li2023terahertz}.
They host an astonishing diversity of physical phenomena such as weak ferromagnetism due to the Dzyaloshinskii-Moriya interaction~\cite{sasani2022origin,bousquet2016non}, magnetoelectricity~\cite{tokunaga2008magnetic,ivanov2023metamagnetic,ivanov2023observation}, multiferroicity~\cite{pyatakov2012magnetoelectric}, electromagnons~\cite{stanislavchuk2016magnon,stanislavchuk2017far}, altermagnetism~\cite{smejkal2022emerging,kimel2024optical,naka2025altermagnetic,song2025altermagnets}, and many others~\cite{li2025recent}.
Orthoferrites hold a particular place as a platform for experimental research in the fields of ultrafast magnetism~\cite{kimel2004laser,kimel2009inertia,gareev2024optical,khokhlov2024double,leenders2024canted}, nonlinear phononics~\cite{juraschek2017ultrafast}, magnonics~\cite{baierl2016nonlinear,zhang2024upconversion,zhang2024coupling,zhang2024spin} and magnetophononics~\cite{nova2017effective,afanasiev2021ultrafast}.
One of the unique magnetic properties of the rare-earth orthoferrites is their tendency to change one of three symmetry allowed magnetic structures for the iron subsystem labeled by $\Gamma_{1}$, $\Gamma_{2}$, and $\Gamma_{4}$ as a result of spin-reorientation transitions under the magnetic interaction between spins of iron $\mathrm{Fe}^{3+}$ and rare-earth $\mathrm{R}^{3+}$ subsystems~\cite{li2023terahertz}.
The most common spin-reorientation transition in orthoferrites is between $\Gamma_{2}$ and $\Gamma_{4}$ magnetic structures through the intermediate state $\Gamma_{24}$~\cite{zhao2016spin,li2023terahertz,li2025recent}.

Many papers focus on the study of magnetic~\cite{shapiro1974neutron,white1982light,koshizuka1980inelastic,venugopalan1985magnetic,wang2025continuous} and lattice~\cite{weber2022emerging,panchwanee2019temperature} dynamics evolution during spin-reorientation transition in orthoferrites in order to reveal new effective mechanisms of interaction between iron and rare-earth subsystems.
In this paper, we report the results of the experimental study on the lattice and magnetic dynamics in the Brillouin zone center of the rare-earth orthoferrite \STFO{} single crystal employing polarized Raman scattering and infrared reflectivity spectroscopic techniques.
The obtained experimental results were supported by the corresponding first-principles lattice dynamics calculations. 
Furthermore, we explored the Raman spectra of magnon modes for all main polarization configurations in magnetic structures $\Gamma_{2}$ and $\Gamma_{4}$ and revealed the temperature evolution of corresponding modes in the range of spin-reorientation transitions.
The interest in this particular compound arises due to the fact that pure \SFO{} and \TFO{} crystals have a spin-reorientation transition $\Gamma_{4} \xleftrightarrow{} \Gamma_{24} \xleftrightarrow{} \Gamma_{2}$ between the most common magnetic structures for orthoferrites at very different temperatures, namely, notably above 300\,K and below 10\,K, respectively~\cite{wu2017crystal,li2023terahertz}.
As a consequence, the spin-reorientation transition in mixed orthoferrite \STFO{} may vary in a very broad temperature range, and, in particular, in the vicinity of the room temperature.
This property opens potential possibilities for developing temperature-controllable magnetic devices.

\section{Materials and Methods}

\subsection{Samples}
The high-quality single crystals of orthoferrite \STFO{} were grown by the four-mirror floating zone technique, as described in detail in Ref.~\cite{wang2015single}.
The XRD oriented single crystals were cut into samples with a normal of the surface along the three main crystallographic axes and polished to optical surface quality.
The samples have a typical thickness of about 1\,mm and a surface size of about $8\times{8}$\,mm$^{2}$.

\subsection{Infrared spectroscopy}
Infrared reflectivity spectra were measured at room temperature using a Bruker IFS 66v/S spectrometer.
The measurements were performed with near-normal incidence (approximately 13$^{\circ}$ from the surface normal).
The spectral ranges of 50--450\,cm$^{-1}$ and 450--5000\,cm$^{-1}$ were covered using DTGS and DLaTGS detectors, respectively, with a resolution of 4\,cm$^{-1}$, 128 scans, and a scanning velocity of 2.2\,kHz.
The linear polarization of light emitted by the globar source was set by the THz linear thin film polarizer along the main crystallographic axes of the samples.
The reflectivity spectra were normalized to the gold mirror reference to obtain absolute values.

\subsection{Raman spectroscopy}
The polarized Raman spectra were measured using the Horiba LabRAM HREvo UV-Vis-NIR-Open spectrometer with the 1800 lines/mm grating equipped with a confocal microscope and a liquid nitrogen cooled CCD detector.
For excitation, a 632.8\,nm line of a HeNe laser (Newport N-LHP 928) was used with a power of 12\,mW.
\textcolor{newtext}{The sample temperature was varied in the range from 78 to 400\,K using a Linkam THMS600 thermal stage.}
Experiments were performed in the backscattering geometry using a Leica PL FLUOTAR $50\times$ (NA = 0.55) long working-distance objective lens to focus the incident beam into a spot of 2\,$\mu$m diameter and to collect the scattered light.
For measurement at ambient conditions, an Olympus MPLN 100$\times$ objective was used to focus the excitation beam into a spot with a diameter of $<1$ $\mu$m.

\subsection{Lattice dynamics calculations}
We have supplemented the experimental results by the lattice dynamics calculations performed in the framework of density functional theory (DFT) with B3LYP hybrid functional~\cite{becke1993density} implemented on \textsc{CRYSTAL14} package~\cite{dovesi2014crystal14}.
The quasi-relativistic ECP$n$MWB pseudo-potentials for $\mathrm{Tb}^{3+}$ ($n = 54$) and $\mathrm{Sm}^{3+}$ ($n = 51$) were used to describe the core electrons including the $4f$ shell~\cite{dolg1989energy,dolg1993combination}.
To describe the valence electrons of rare-earth ions, the ECP$n$MWB-I basis sets were used~\cite{dolg1989energy,yang2005valence}.
All electron basis sets have been used to describe ions with a contraction scheme of (842111s)-(6311p)-(411d)-(1f) for $\mathrm{Fe}$, and (6211s)-(411p)-(1d) for $\mathrm{O}$~\cite{peintinger2013consistent}.
The putative ferromagnetic spin configuration was used in the calculations.
The reciprocal space was sampled by $8\times8\times8$ $k$-point mesh.
The parameters that establish the accuracy in evaluating the Coulomb and Hartree-Fock exchange series were set as 8, 8, 8, 8, and 16~\cite{dovesi2014crystal14}.
The threshold on the self-consistent field energy was set to $10^{-9}$ Hartree for geometry optimization and $10^{-8}$ Hartree for frequency calculation.
The phonon spectra of \TFO{} and \SFO{} were calculated in harmonic approximation in the center of the Brillouin zone by means of numerical second derivatives of the total energy.
The static dielectric tensors, Born effective charges, and phonon intensities were calculated using the CPHF/KS approach~\cite{maschio2012infrared,maschio2013raman}.
It was shown that the phonon frequencies in rare-earth orthoferrites depend linearly on the ionic radius of the $\it{R}$ ion~\cite{weber2016raman}.
From this, the phonon frequencies for \STFO{} were estimated using those for \SFO{} and \TFO{} and taking into account the effective ionic radii of the $\mathrm{R}$ ion as $r_{\it{R}} = 0.55\,r_{\mathrm{Sm}}(1.079\,\mathrm{\AA}) + 0.45\,r_{\mathrm{Tb}}(1.040\,\mathrm{\AA}) \simeq 1.0615$\,\AA.
The obtained phonon frequencies for \SFO{}, \TFO{}, and \STFO{} are listed in Table~\ref{tab:dft_phonons}.

\section{Results and Discussions}

\subsection{Crystal and magnetic structures and phonons}

\begin{figure}
\centering
\includegraphics[width=1\columnwidth]{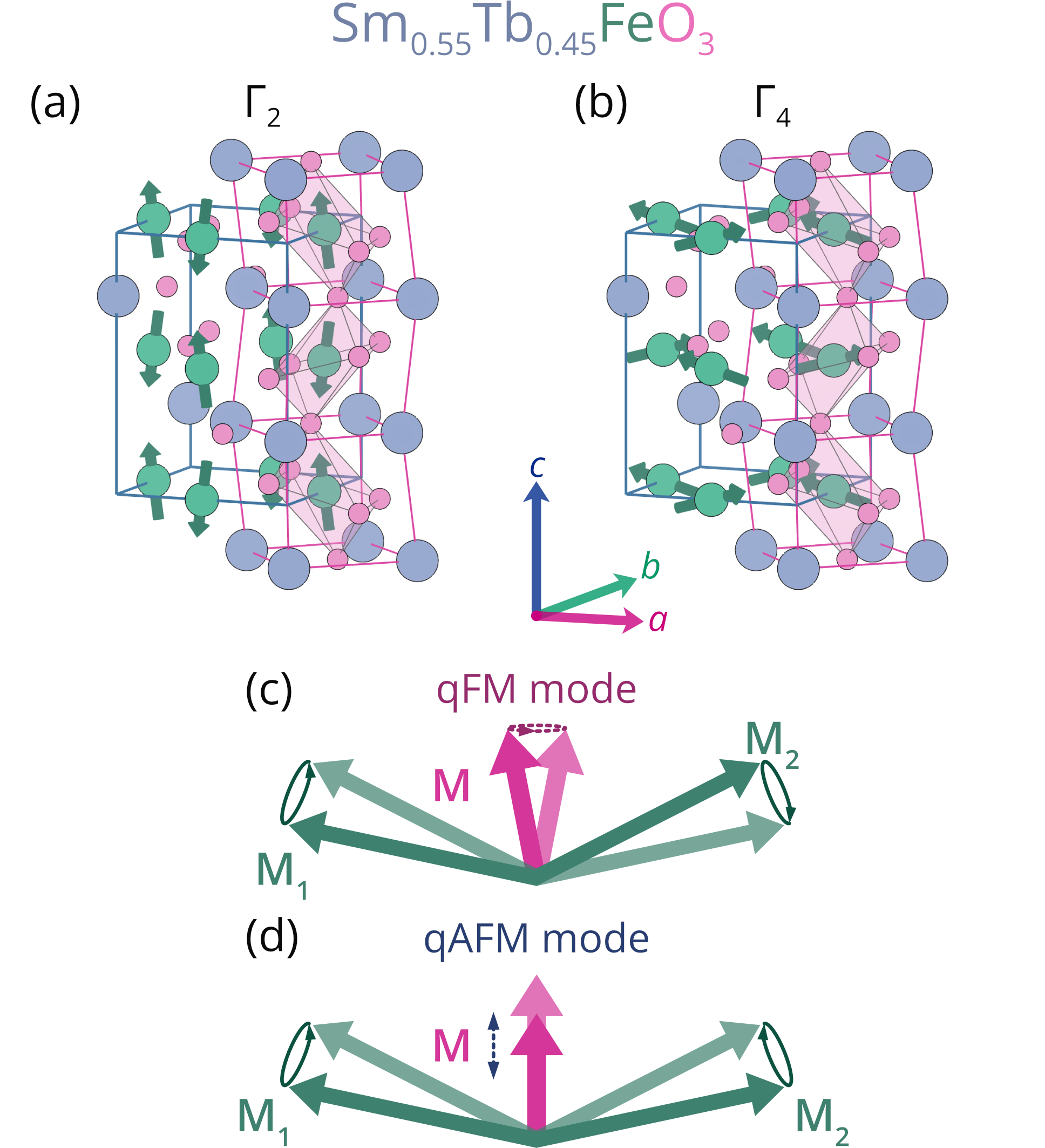}
\caption{\label{fig:structure}
The crystal and magnetic structures of \STFO{} in the $Pbnm$ setting in the (a)~$\Gamma_{2}$ and (b)~$\Gamma_{4}$ magnetic configurations.
(c)~The precession of weak ferromagnetic moment $\mathbf{M} = \mathbf{M}_{1} + \mathbf{M}_{2}$ for quasi-ferromagnetic (qFM) mode and (d) amplitude oscillation of $\mathbf{M}$ for quasi-antiferromagnetic (qAFM) mode in orthoferrites.
}
\end{figure}

\begin{table}
\caption{\label{tab:RFeO3_parameters} Lattice parameters $a$, $b$, and $c$ (in \AA), unit cell volume $V$ (in \AA$^{3}$), static $\varepsilon_{0}$ and high-frequency $\varepsilon_{\infty}$ anisotropic dielectric permittivities in \SFO{}, \TFO{}, and \STFO{} obtained from experiments in comparison with our results of DFT calculations.
}
\begin{ruledtabular}
\begin{tabular}{ccccccc}
                           & \multicolumn{2}{c}{\SFO}                            & \multicolumn{2}{c}{\TFO}                                          & \multicolumn{2}{c}{\STFO} \\ \cmidrule{2-3} \cmidrule{4-5} \cmidrule{6-7}
                           & Exp.\footnote{Ref.~\cite{weber2016raman}.} & DFT    & Exp.\footnote{Ref.~\cite{dubrovin2024lattice}.} & DFT    & Exp.   & DFT    \\ \midrule
$a$                        & 5.400                                      & 5.475  & 5.365                                           & 5.404  & 5.395  & 5.443  \\
$b$                        & 5.584                                      & 5.700  & 5.634                                           & 5.684  & 5.612  & 5.693  \\ 
$c$                        & 7.768                                      & 7.820  & 7.623                                           & 7.755  & 7.688  & 7.791  \\ 
$V$                        & 234.23                                     & 244.04 & 230.39                                          & 238.21 & 232.77 & 241.42 \\ \midrule
$\varepsilon^{a}_{0}$      & ---                                        & 33.331 & 21.7                                            & 31.886 & 25.18  & 32.681 \\
$\varepsilon^{b}_{0}$      & ---                                        & 30.96  & 20.5                                            & 29.426 & 23.58  & 30.270 \\
$\varepsilon^{c}_{0}$      & ---                                        & 30.182 & 22.6                                            & 31.081 & 24.86  & 30.587 \\ \midrule
$\varepsilon^{a}_{\infty}$ & ---                                        & 5.215  & 4.81                                            & 5.141  & 5.4    & 5.182  \\ 
$\varepsilon^{b}_{\infty}$ & ---                                        & 5.237  & 4.82                                            & 5.203  & 5.39   & 5.222  \\
$\varepsilon^{c}_{\infty}$ & ---                                        & 5.094  & 4.79                                            & 5.017  & 5.36   & 5.059  \\ 
\end{tabular}
\end{ruledtabular}
\end{table}

The rare-earth orthoferrite \STFO{} has the orthorhombically distorted perovskite crystal structure with the centrosymmetric space group $Pbnm$ [No.~62, $D^{16}_{2h}$] and four formula units per unit cell $Z = 4$~\cite{wu2017crystal}.
The $Pbnm$ is an unconventional setting of the $Pnma$ space group, which is conventionally used for orthoferrites~\cite{li2023terahertz}.
The lattice parameters of \STFO{} at room temperature are listed in Table~\ref{tab:RFeO3_parameters}, in comparison with those of \SFO{}, and \TFO{}.
The unit cell contains 20 ions occupying the Wyckoff positions $4c$ for $\mathrm{Sm}^{3+}$ and $\mathrm{Tb}^{3+}$, $4b$ for $\mathrm{Fe}^{3+}$, and $4c$ and $8d$ for $\mathrm{O}^{2-}$.

Below the N{\'e}el temperature $T_{N}(\mathrm{Fe})\simeq650$\,K the $\mathrm{Fe}^{3+}$ spins arrangement is a canted antiferromagnetic structure with antiparallel ordering along the $a$ axis ($G_{a}$) and a weak ferromagnetic moment along the $c$ axis ($F_{c}$) due to the Dzyaloshinskii-Moriya interaction, which corresponds to the $\Gamma_{4}(G_{a}F_{c})$ magnetic structure in Bertaut's notation.
\SFO{}, \STFO{} and \TFO{} exhibit one of the most common spin-reorientation transitions in orthoferrites in which the $\mathrm{Fe}^{3+}$ spins are rotated continuously in the $ac$ plane in the temperature range from $T_{1}$ to $T_{2}$ ($T_{2} < T_{1} < T_{N}$) from the $\Gamma_{4}(G_{a}F_{c})$ [Fig.~\ref{fig:structure}(b)] to $\Gamma_{2}(G_{c}F_{a})$ [Fig.~\ref{fig:structure}(a)] through the intermediate $\Gamma_{24}(G_{ac}F_{ac})$ state~\cite{wu2017crystal,yamaguchi1974theory}.
The highest spin-reorientation transition temperatures $T_{1} \simeq 480$\,K and $T_{2} \simeq 450$\,K are observed in $\mathrm{SmFeO}_{3}$~\cite{weber2022emerging}.
Isovalent substitution in $\mathrm{Sm}_{1-x}\mathrm{R}_{x}\mathrm{FeO}_{3}$ reduces the transition temperatures $T_{1}$ and $T_{2}$ to the required values by varying the concentration of $\it{R}^{3+}$ ions~\cite{wang2015single,wu2017crystal,li2025recent}.


The group-theoretical analysis of $Pbnm$ orthoferrites \RFO{} predicts 60 phonons at the center of the Brillouin zone~\cite{kroumova2003bilbao}:
\begin{equation}
\label{eq:group_irrep_total_orth}
\begin{gathered}
\Gamma_{\mathrm{total}} =
\underbrace{B_{1u}(z) \oplus B_{2u}(y) \oplus B_{3u}(x)}_{\Gamma_{\mathrm{acoustic}}} \\ \oplus 
\underbrace{7 B_{1u}(z) \oplus 9 B_{2u}(y) \oplus 9 B_{3u}(x)}_{\Gamma_{\mathrm{IR}}} \\ \oplus
\underbrace{7 A_{g}(x^{2},y^{2},z^{2}) \oplus 7 B_{1g}(xy) \oplus 5 B_{2g}(xz) \oplus 5 B_{3g}(yz)}_{\Gamma_{\mathrm{Raman}}} \\ \oplus
\underbrace{8A_{u}(xyz)}_{\Gamma_{\mathrm{silent}}},
\end{gathered}
\end{equation}
among which there are 3 acoustic, 25 infrared-active, 24 Raman-active, and 8 silent modes.
The basis functions are given in parentheses.

\subsection{Infrared-active phonons}

\begin{figure*}
\centering
\includegraphics[width=2\columnwidth]{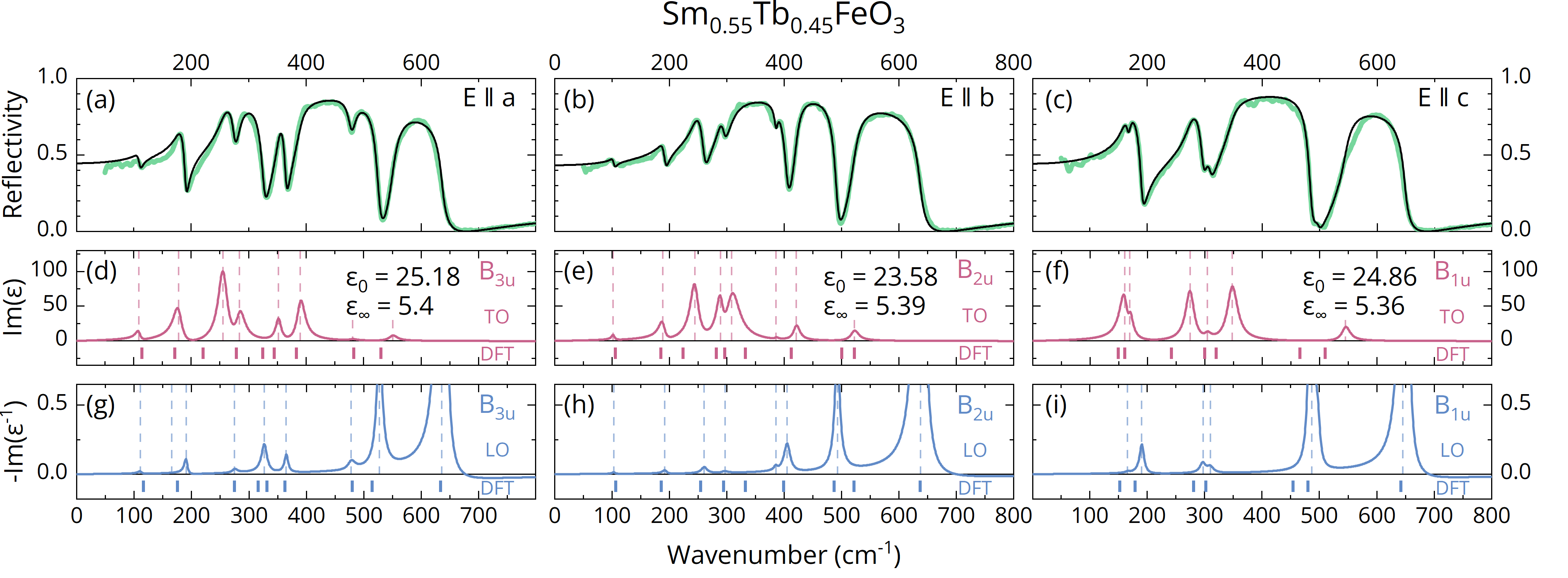}
\caption{\label{fig:reflectivity}
The infrared reflectivity spectra with the electric field $\mathbf{E}$ of radiation polarized along the (a)~$a$, (b)~$b$ and $c$~(c) axes of the \STFO{} at room temperature in the $Pbnm$ setting. 
The solid black lines are fits using Eq.~\eqref{eq:reflectivity} with a complex dielectric permittivity $\varepsilon = \varepsilon_{1} - i\varepsilon_{2}$ from Eq.~\eqref{eq:epsilon_TOLO}.
Spectra of the $\Im[\varepsilon(\omega)]$ and $-\Im[\varepsilon^{-1}(\omega)]$ from fits which correspond to the TO and LO infrared-active phonons with (d), (g)~$B_{3u}$, (e), (h)~$B_{2u}$, and (f), (i)~$B_{1u}$ symmetries, respectively.
Vertical dashed gray lines denote the experimental frequencies of infrared-active phonons.
Color sticks at the bottom of plots represent the calculated phonon frequencies.
}
\end{figure*}

\begin{table}
\caption{\label{tab:IR_phonons} Experimental frequencies $\omega$ (cm$^{-1}$), dampings $\gamma$ (cm$^{-1}$), and dielectric strengths $\Delta\varepsilon$ of the TO and LO polar phonons in \STFO{} at room temperature in comparison with the results of DFT calculations presented in parentheses.}
\begin{ruledtabular}
\begin{tabular}{cccccccccccccc}
Sym. & & \multicolumn{2}{c}{$\omega_{\mathrm{TO}}$} & & $\gamma_{\mathrm{TO}}$ & & \multicolumn{2}{c}{$\omega_{\mathrm{LO}}$} &  & $\gamma_{\mathrm{LO}}$ & & \multicolumn{2}{c}{$\Delta\varepsilon$} \\ \midrule
\multirow{7}{*}{$B_{1u}$} & & 160.4 & (149.4) & & 18.3 & & 166.9 & (152.5) & & 7.3 & & 7.14 & (3.84)\\
                          & & 169.1 & (160.8) & & 9.1 & & 191.2 & (179.7) & & 9.4 & & 1.68 & (6.39)\\
                          & & 274.5 & (242.5) & & 18.2 & & 297 & (281.4) & & 11.7 & & 4.74 & (9.30)\\
                          & & 304.8 & (300.3) & & 13.8 & & 310 & (302.8) & & 14.3 & & 0.34 & (0.52)\\
                          & & 347.8 & (320.2) & & 22 & & 484.8 & (454.1) & & 10.2 & & 4.95 & (4.75)\\
                          & & 495.6 & (465.8) & &  14 & &  498 & (480.1) & & 11.1  & & 0.01 & (0.11)\\
                          & & 545.5 & (510.1) & & 17.8 & & 647.8 & (641.9) & & 9.8 & & 0.66 & (0.60)\\ \midrule
\multirow{9}{*}{$B_{2u}$} & & 102.3 & (106.3) & & 6.9 & & 103.2 & (106.5) & & 6.4 & & 0.45 & (0.13) \\
                        &  & 188.6 & (185.2) & & 13.3 & & 192.5 & (185.7) & & 9.1 & & 1.56 & (0.37)\\
                        &  & 244.6 & (223.7) & & 17.9 & & 260.6 & (254.5) & & 12.4 & & 5.77 & (14.92)\\
                        &  & 289.2 & (282.3) & & 14.2 & & 295.5 & (294.6) & & 11.6 & & 2.87 & (7.49)\\
                        &  & 308.7 & (296.4) & & 29.1 & & 385.5 & (332.7) & & 6.7 & & 6.29 & (1.18)\\
                         & & 386.3 & (332.7) & &  6.7 & & 405.4 & (399.1) & & 13 & & 0.04 & (0.00)\\
                         & & 421.5 & (412.7) & & 13.6 & & 492.6 & (487) & & 11 & & 0.69 & (0.64)\\
                         & &  --- & (500.7)  & & ---  &  & --- & (521.9)  & & --- & & --- & (0.26) \\
                         & & 522.5 & (522.7) & & 17.1  & & 636.6 & (637) & &  12 & & 0.49 & (0.02)\\ \midrule
\multirow{9}{*}{$B_{3u}$} & & 109.3 & (114.9) & & 10.2 & &  111 & (116.3) & &  7.8 & & 0.95 & (1.04) \\
                          & & 178.7 & (171.7) & & 18.7 & & 190.8 & (176) & &  6.8 & & 4.55 & (3.18)\\ 
                          & & 255.8 & (220.9) & & 19.8 & & 275.3 & (274.7) & & 11.4 & & 7.86 & (19.34)\\ 
                          & & 284.4 & (278.8) & & 18.4 & & 326.9 & (316.6) & & 11.8 & & 2.48 & (0.73)\\ 
                          & &  --- & (324.9)  & & ---  & & --- &  (331.5) & & ---  & & --- & (0.27) \\ 
                          & & 352.1 & (344.5) & & 11.7 & & 365.1 & (362.7) & & 7.9 & & 1.05 & (0.91)\\ 
                          & & 390.2 & (383.1) & & 17.5  & & 478.3 & (480.2) & & 15.2 & &  2.6 & (1.79)\\ 
                          & & 480.9 & (483.2) & & 16.1  & & 527.3 & (514.9) & & 11.8 & & 0.06 & (0.05)\\ 
                         & & 550.7 & (530.5) & & 17.5 & & 636.1 & (633.8) & & 10.2 &  & 0.25 & (0.20)\\
\end{tabular}
\end{ruledtabular}
\end{table}

The infrared reflectivity spectra measured for the electric field $\mathbf{E}$ of radiation polarized along the $a$, $b$, and $c$ axes of \STFO{} at room temperature are shown by the green lines in Figs.~\ref{fig:reflectivity}(a)--\ref{fig:reflectivity}(c), respectively.
The reflection bands in the spectra correspond to the $B_{1u}$, $B_{2u}$, and $B_{3u}$ phonons, which are active for specific polarizations according to Eq.~\eqref{eq:group_irrep_total_orth}.
These reflectivity spectra were fitted by using the factorized form of the complex dielectric permittivity~\cite{gervais1974anharmonicity}
\begin{equation}
\label{eq:epsilon_TOLO}
\varepsilon(\omega) = \varepsilon_{1}(\omega) - i\varepsilon_{2}(\omega) = \varepsilon_{\infty}\prod\limits_{j}\frac{{\omega^{2}_{j\textrm{LO}}} - {\omega}^2 + i\gamma_{j\textrm{LO}}\omega}{{\omega^{2}_{j\textrm{TO}}} - {\omega}^2 + i\gamma_{j\textrm{TO}}\omega},
\end{equation}
where $\varepsilon_{\infty}$ is the high-frequency dielectric permittivity, $\omega_{j\textrm{TO}}$, $\omega_{j\textrm{LO}}$, $\gamma_{j\textrm{TO}}$, and $\gamma_{j\textrm{LO}}$ correspond to $\textrm{TO}$ and $\textrm{LO}$ frequencies ($\omega_{j}$) and dampings ($\gamma_{j}$) of the $j$th polar phonon of the specific symmetry, respectively.
Multiplication occurs over all polar phonons with the specific symmetry, which are active for this polarization of the incident light.
We note that Eq.~\eqref{eq:epsilon_TOLO} at $\omega = 0$ converges to the well-known Lyddane-Sachs-Teller relation~\cite{lyddane1941polar}. 
For normal incidence, the infrared reflectivity $R(\omega)$ and complex dielectric function $\varepsilon(\omega)$ are related to each other via the Fresnel equation~\cite{born2013principles}
\begin{equation}
\label{eq:reflectivity}
R(\omega) = \Bigl|\frac{\sqrt{\varepsilon(\omega)} - 1}{\sqrt{\varepsilon(\omega)} + 1}\Bigr|^2.
\end{equation}
There is a fair agreement between the reflectivity spectra (green lines) and the fits (black lines) from Eqs.~\eqref{eq:epsilon_TOLO} and~\eqref{eq:reflectivity} as can be seen in Figs.~\ref{fig:reflectivity}(a)--\ref{fig:reflectivity}(c).
The peaks in the spectra of the imaginary part of the dielectric permittivity $\Im[\varepsilon(\omega)]$ [Figs.~\ref{fig:reflectivity}(d)--\ref{fig:reflectivity}(f)] and the inverse dielectric permittivity $\Im[\varepsilon^{-1}(\omega)]$ [Figs.~\ref{fig:reflectivity}(g)--\ref{fig:reflectivity}(i)] correspond to the frequencies of transverse (TO) and longitudinal (LO) polar phonons, respectively~\cite{schubert2004infrared}.
The parameters of the $B_{1u}$, $B_{2u}$, and $B_{3u}$ infrared-active phonons derived from the fits are listed in Table~\ref{tab:IR_phonons}. 
The obtained results are close to those reported in the literature for other orthoferrites~\cite{komandin2023electric,gomes2024lattice,dubrovin2024lattice}.

The static dielectric permittivity $\varepsilon_{0} = \varepsilon_{\infty} + \sum_{j}\Delta\varepsilon_{j}$ is determined by the infrared-active phonons via their dielectric strengths~\cite{gervais1983long}
\begin{equation}
\label{eq:oscillator_strength_TOLO}
\Delta\varepsilon_{j}  =  \frac{\varepsilon_{\infty}}{{\omega^{2}_{j\textrm{TO}}}}\frac{\prod\limits_{k}{\omega^{2}_{k\textrm{LO}}}-{\omega^{2}_{j\textrm{TO}}}}{\prod\limits_{k\neq{}j}{\omega^{2}_{k\textrm{TO}}}-{\omega^{2}_{j\textrm{TO}}}}.
\end{equation}
The dielectric strengths $\Delta\varepsilon$ evaluated using Eq.~\eqref{eq:oscillator_strength_TOLO} are listed in Table~\ref{tab:IR_phonons}.
The anisotropic static $\varepsilon_{0}$ and high frequency $\varepsilon_{\infty}$ dielectric permittivities obtained from fits are given in Figs.~\ref{fig:reflectivity}(a)--\ref{fig:reflectivity}(c).
This value of $\varepsilon_{0}$ for \STFO{} is in fair agreement with the data reported in the literature on several orthoferrites~\cite{balbashov1995submillimeter,stanislavchuk2016magnon,komandin2023electric}.

\subsection{Raman-active phonons}

\begin{figure*}
\centering
\includegraphics[width=2\columnwidth] {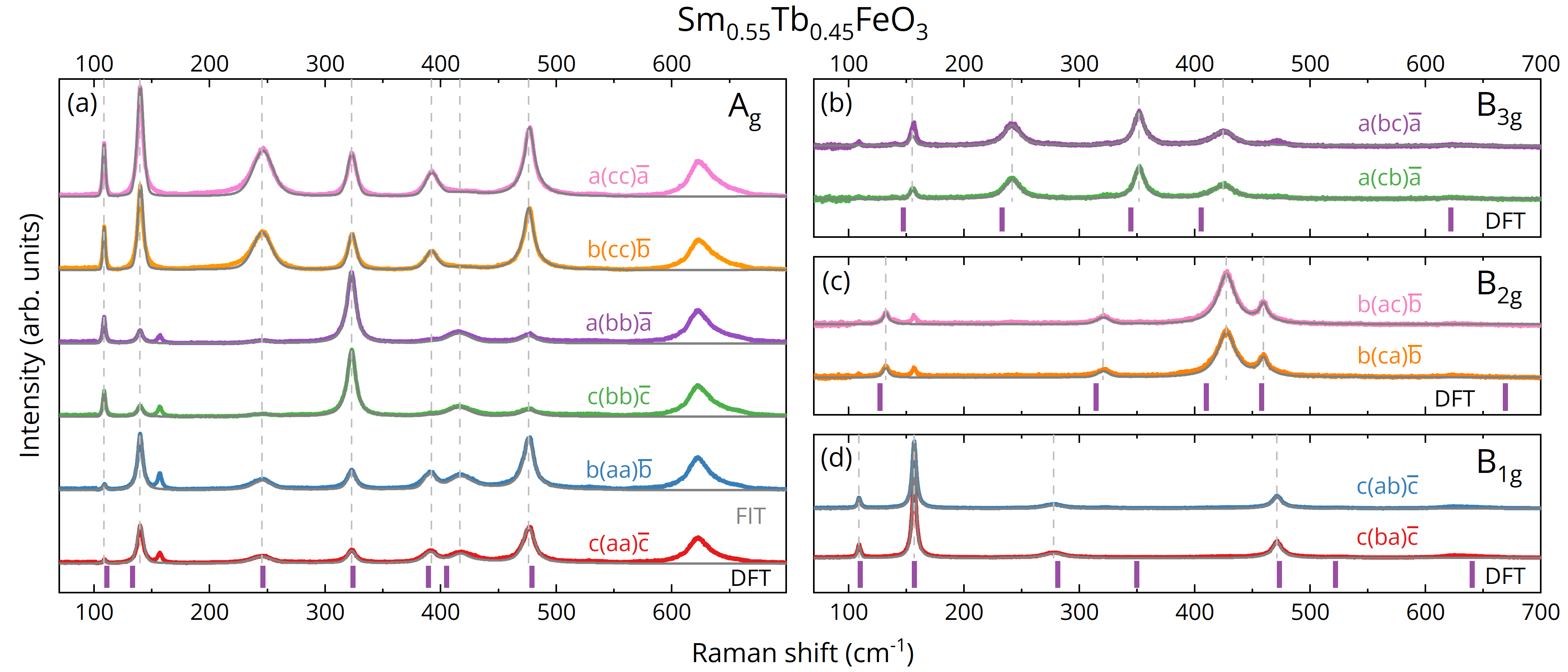}
\caption{\label{fig:raman}
The polarized Raman spectra of the (a)~$A_{g}$, (b)~$B_{3g}$, (c)~$B_{2g}$, and (d)~$B_{1g}$ phonons for \STFO{} at ambient conditions.
The polarization configurations are given in Porto's notation as described in the text in the $Pbnm$ setting.
The solid gray lines are fits of the phonons active in the corresponding polarization configuration as described in the text.
The frequencies of these phonons are denoted by the vertical dashed lines.
Purple sticks at the bottom of each plot represent the calculated phonon frequencies.
}
\end{figure*}

The intensity of the light scattered by a Raman-active phonon is determined by the following relation~\cite{loudon2001raman}:
\begin{equation}
\label{eq:raman}
I \propto |\bm{e}_{s} \mathcal{R} \bm{e}_{i}|^2,
\end{equation}
where $\mathcal{R}$ is the Raman tensor, $\bm{e}_{s}$ and $\bm{e}_{i}$ are the unit vectors of the polarization for the incident and scattered light in the crystallographic axes coordinate system, respectively.
The Raman tensors of Raman-active phonons for $Pbnm$ orthoferrites are defined as~\cite{martin2001melting,kroumova2003bilbao} 
\begin{equation}
    \label{eq:raman_tensors}
    \begin{gathered}
        \mathcal{R}_{A_{g}} =
        \begin{pmatrix}
             a_{1} & 0     & 0 \\ 
             0     & a_{2} & 0 \\ 
             0     &     0 & a_{3}
        \end{pmatrix}
        ,\quad
        \mathcal{R}_{B_{1g}} = 
        \begin{pmatrix}
             0     & a_{4} & 0 \\ 
             a_{4} &     0 & 0 \\
             0     &     0 & 0
        \end{pmatrix},\\[1ex]
        \mathcal{R}_{B_{2g}} = 
        \begin{pmatrix}
             0     & 0 & a_{5} \\ 
             0     & 0 & 0     \\
             a_{5} & 0 & 0
        \end{pmatrix}
        ,\quad
        \mathcal{R}_{B_{3g}} = 
        \begin{pmatrix}
             0 & 0     & 0     \\ 
             0 & 0     & a_{6} \\
             0 & a_{6} & 0
        \end{pmatrix},
    \end{gathered}
\end{equation}
where $a_{1}$--$a_{6}$ are Raman tensor elements.
Thus, Raman-active phonons of a given symmetry can be selectively distinguished by using specific polarization configurations of incident and scattered light with respect to the main crystallographic axes.
The polarization configuration is usually given in Porto's notation, according to which a set of four symbols is used, $\bm{k}_{i} (\bm{e}_{i}\bm{e}_{s}) \bm{k}_{s}$, where $\bm{k}_{i}$ and $\bm{k}_{s}$ are the directions of the propagation of the incident and scattered light (for the back-scattering geometry these directions are opposite $\bm{k}_{i}=\overline{\bm{k}}_{s}$)~\cite{damen1966raman}.
According to Eqs.~\eqref{eq:raman_tensors} and~\eqref{eq:raman}, the $A_{g}$ phonons are active for the parallel polarization settings $\bm{e}_{i}\parallel{\bm{e}_{s}}$ 
whereas $B_{1g}$, $B_{2g}$, and $B_{3g}$ phonons are distinguishable in the crossed configurations $\bm{e}_{i}\perp{\bm{e}_{s}}$.

\begin{table}
\caption{\label{tab:raman_phonons} Frequencies (cm$^{-1}$), full widths at half-maximum (FWHM, cm$^{-1}$), and maximal intensity (counts/s) of the Raman-active $A_{g}$, $B_{1g}$, $B_{2g}$, and $B_{3g}$ phonons for \STFO{} at ambient conditions.
The calculated phonon frequencies are given in parentheses.}
\begin{ruledtabular}
\begin{tabular}{cccccccc}
Sym. & & \multicolumn{2}{c}{Freq.} & & FWHM & & Intensity \\ \midrule
\multirow{7}{*}{$A_{g}$} & & 108.7 & (111) & & 2.9 & & 35.4 \\
                         & & 139.9 & (133.1) & & 6.1 & & 74.3 \\
                         & & 245.5 & (245.9) & & 21.2 & & 31.1 \\
                         & & 322.8 & (324.1) & & 9.1 & & 47.9 \\
                         & & 392.1 & (389.3) & & 12.6 & & 14.9 \\
                         & & 416.7 & (405.1) & & 27.7 & & 9.3 \\ 
                         & & 476.2 & (478.9) & & 13.2 & & 46.9 \\ \midrule
\multirow{7}{*}{$B_{1g}$} & & 109.1 & (109.9) & & 3.4 & & 7.5 \\
                         & & 156.9 & (157) & & 4.5 & & 49.5 \\
                         & & 277.9 & (281.4) & & 20.4& & 2.9 \\
                         & & ---   & (349.6) & & --- & & --- \\
                         & & 471.2 & (473.3) & & 9.6  & & 9.4 \\
                         & & ---   & (521.9)& & --- & & --- \\
                         & & ---   & (640.4)& & --- & & --- \\ \midrule
\multirow{5}{*}{$B_{2g}$} & & 132.3 & (127.3) & & 5.6 & & 2.1 \\
                         & & 320.8 & (314.6) & & 12.9 & & 1.1 \\
                         & & 427.5 & (409.9) & & 18.4 & & 8.5 \\
                         & & 459.5 & (458) & & 9.4 &  & 3.2 \\ 
                         & & ---   & (669.1) & & --- & & --- \\ \midrule
\multirow{5}{*}{$B_{3g}$} & & 155.4 & (147.5) & & 5.6 & & 1.6\\ 
                          & & 241.8 & (233) & & 8.1 & & 3.7 \\
                          & & 351.9 & (344.6) & & 5.6 & & 5.8 \\
                          & & 424.6 & (405.6) & & 11.9 & & 2.5 \\
                          & & ---   & (621.9) & & --- & & --- \\
\end{tabular}
\end{ruledtabular}
\end{table}

Figure~\ref{fig:raman} shows the Raman polarized spectra of \STFO{} at ambient conditions.
These spectra were analyzed and $A_{g}$, $B_{1g}$, $B_{2g}$, and $B_{3g}$ Raman-active phonons were reliably identified in parallel and crossed polarizations, respectively.
The small leakage of phonons in forbidden polarizations was observed presumably due to the slight misalignment of the polarization of light with respect to the crystallographic axes in the samples and the almost unavoidable depolarization effect in the optical elements.
The frequencies, full widths at half maximum (FWHMs), and intensities of the Raman-active phonons were extracted by fitting the Raman spectra with a sum of Voigt profiles implemented in \textsc{fityk} software~\cite{wojdyr2010fityk} without Bose correction.
These results are listed in Table~\ref{tab:raman_phonons}.
The frequencies of the Raman-active phonons are close to those for other orthoferrites~\cite{koshizuka1980inelastic,white1982light,venugopalan1985magnetic,dubrovin2024lattice}.

The observed strong asymmetric line with a frequency ${\simeq}\,630$\,cm$^{-1}$ appearing in parallel polarizations only was previously reported in \SFO~\cite{weber2016raman,gupta2018temperature,panchwanee2019temperature,warshi2020cluster} and other orthoferrites~\cite{venugopalan1983raman,gupta2002lattice,weber2016raman,coutinho2017structural,raut2018grain,nagrare2018hyperfine,gupta2018temperature,ponosov2020lattice}.
According to Refs.~\cite{weber2016raman,ponosov2020lattice}, the frequency and polarization selection rules of this line seem to be independent of the $\it{R}$ ions, including $\mathrm{Y}$.
Moreover, the calculated frequencies of the $A_{g}$ phonons are lower than this excitation~\cite{gupta2002lattice,weber2016raman,dubrovin2024lattice}.
This allows us to assume that this line is not an $A_{g}$ Raman-active phonon, but it is related to the chemical defects in the lattice.
However, it is worth noting that intensity and spectra depend on the $\it{R}$ ion and this dependence is reproduced for different orthoferrite crystals of the same composition.
While this line exhibits low intensity in \TFO~\cite{venugopalan1985magnetic,weber2016raman,dubrovin2024lattice}, its intensity is markedly greater in \SFO~\cite{weber2016raman} and in \STFO{}.

\subsection{Magnons}

There are four $\mathrm{Fe}^{3+}$ ions in the orthoferrite unit cell, which gives two doubly degenerate magnon branches called acoustic and exchange.
The Dzyaloshinskii-Moriya interaction leads to a net magnetic moment due to a small spin canting (${\simeq}\,8.5$\,mrad) and splitting of the magnon branches~\cite{li2023terahertz}.
The acoustic branch is split on the quasi-ferromagnetic (qFM, ${\simeq}\,10$\,cm$^{-1}$) and quasi-antiferromagnetic (qAFM, ${\simeq}\,20$\,cm$^{-1}$) magnon modes.
These modes can be represented as the precession of the ferromagnetic vector $\mathbf{M} = \mathbf{M}_{\mathrm{1}} + \mathbf{M}_{\mathrm{2}}$ for qFM [see Fig.~\ref{fig:structure}(c)], and by amplitude oscillation of $\mathbf{M}$, which corresponds to the precession of the antiferromagnetic vector $\mathbf{L} = \mathbf{M}_{\mathrm{1}} - \mathbf{M}_{\mathrm{2}}$ for qAFM [see Fig.~\ref{fig:structure}(d)] magnon modes, where $\mathbf{M}_{\mathrm{1}}$ and $\mathbf{M}_{\mathrm{2}}$ are magnetizations of the oppositely directed sublattices.
The exchange magnon branch (${\simeq}\,450$\,cm$^{-1}$) is also split by the Dzyaloshinskii-Moriya interaction.
This magnon mode interacts extremely weakly with light, but appears in inelastic neutron scattering experiments~\cite{shapiro1974neutron}.
However, the neutron experiments do not allow resolving the splitting of acoustic and exchange magnon branches.

The qFM and qAFM magnon modes in orthoferrites are clearly observable in Raman scattering~\cite{white1982light,venugopalan1983raman,venugopalan1985magnetic,koshizuka1988raman}, pump-probe~\cite{kimel2005ultrafast,gareev2024optical} and THz~\cite{balbashov1995submillimeter,stanislavchuk2016magnon,stanislavchuk2017far,li2023terahertz,komandin2023electric} experiments.
The Raman tensors for qFM and qAFM magnon modes for orthoferrites in the $Pbnm$ setting are given for low-temperature $\Gamma_{2}$ magnetic structure~\cite{baryakhtar1983light} 
\begin{equation}
    \label{eq:raman_tensors_magnon_Gamma2}
    \begin{gathered}
        \mathcal{R}_{\mathrm{qFM}}^{\Gamma_{2}}(B_{g}) = 
        \begin{pmatrix}
             0     & b_{1} & b_{2} \\ 
             b_{3} & 0     & 0     \\
             b_{4} & 0     & 0
        \end{pmatrix}
        ,\\ 
        \mathcal{R}_{\mathrm{qAFM}}^{\Gamma_{2}}(A_{g}) = 
        \begin{pmatrix}
             b_{5} & 0     & 0 \\ 
             0     & b_{6} & b_{7} \\
             0     & b_{8} & b_{9}
        \end{pmatrix},
    \end{gathered}
\end{equation}
and for high-temperature $\Gamma_{4}$ magnetic structure~\cite{cracknell1969scattering,white1982light,baryakhtar1983light}
\begin{equation}
    \label{eq:raman_tensors_magnon_Gamma4}
    \begin{gathered}
        \mathcal{R}_{\mathrm{qFM}}^{\Gamma_{4}}(B_{g}) = 
        \begin{pmatrix}
             0     & 0     & c_{1} \\ 
             0     & 0     & c_{2} \\
             c_{3} & c_{4} & 0
        \end{pmatrix}
        ,\\
        \mathcal{R}_{\mathrm{qAFM}}^{\Gamma_{4}}(A_{g}) = 
        \begin{pmatrix}
             c_{5}  & c_{6} & 0     \\ 
             c_{7}  & c_{8} & 0     \\
             0      & 0     & c_{9}
        \end{pmatrix},
    \end{gathered}
\end{equation}
where $b_{1}$--$b_{9}$ and $c_{1}$--$c_{9}$ are Raman tensor elements.

\begin{figure*}
\centering
\includegraphics[width=2\columnwidth]{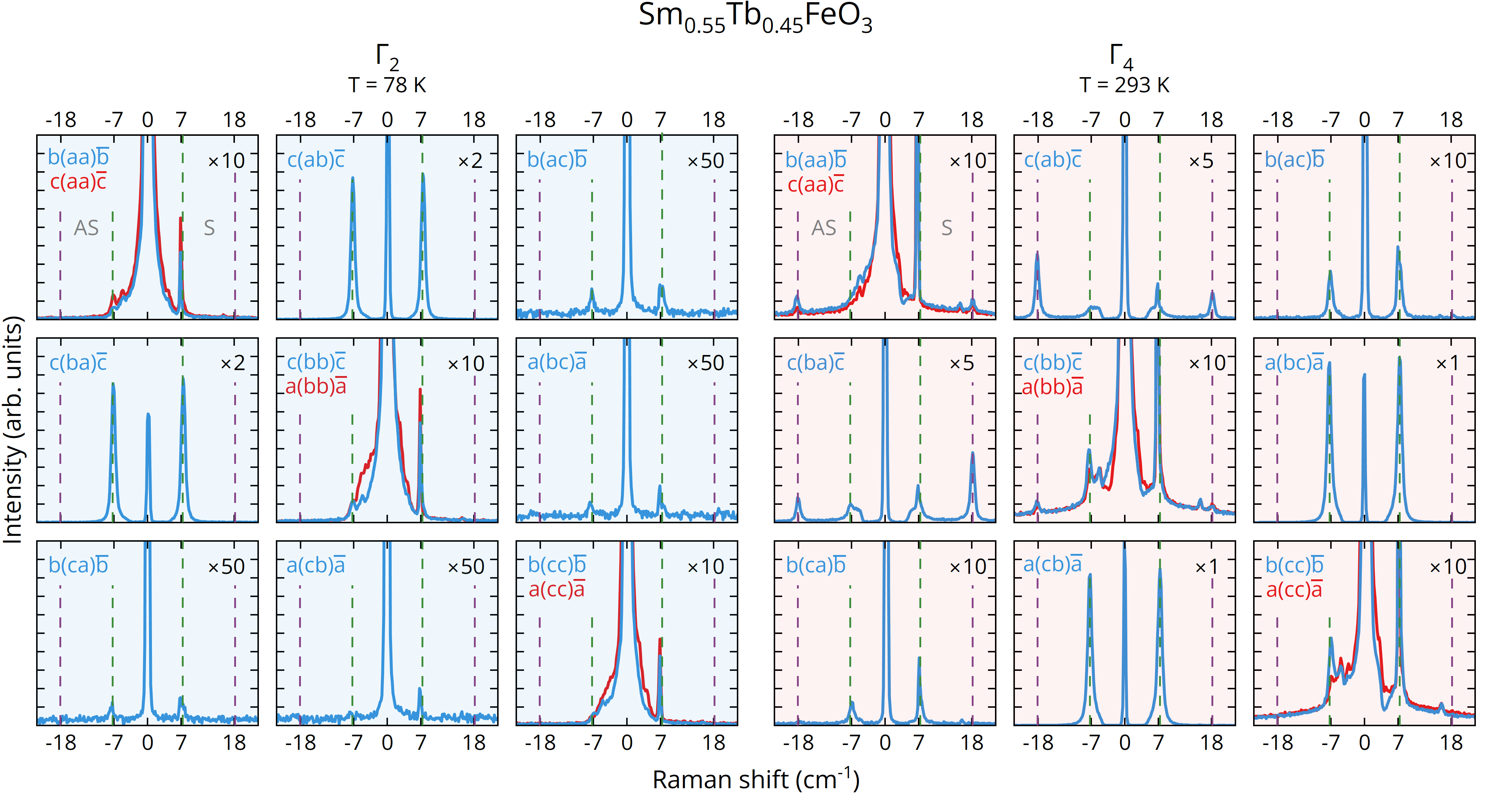}
\caption{\label{fig:magnon}
The polarized Raman spectra of the quasi-ferromagnetic (qFM, $\simeq{\pm7}$\,cm$^{-1}$) and quasi-antiferromagnetic (qAFM, $\simeq{\pm}18$\,cm$^{-1}$) modes in \STFO{} in $\Gamma_{2}$ (blue shaded area, $T = 78$\,K) and $\Gamma_{4}$ (red shaded area, $T = 293$\,K) magnetic structures.
The polarization configurations are given in Porto's notation as described in the text in the $Pbnm$ setting.
The anti-Stokes (AS) and Stokes (S) lines are denoted. 
The intensity of some spectra is multiplied by a factor denoted by the ``$\times$'' sign to highlight the weaker excitations.
}
\end{figure*}

The Raman spectra in the range from -20 to 20\,cm$^{-1}$ for all parallel and crossed polarization configurations along the main crystallographic axes for the $\Gamma_{2}$ at $T = 78$\,K (left panels with blue shaded area) and $\Gamma_{4}$ at 293\,K (right panels with red shaded area) magnetic structures in \STFO{} are shown in Fig.~\ref{fig:magnon}.
The spectra show anti-Stokes and Stokes lines for two excitations at frequencies of $\simeq{\pm}7$\,cm$^{-1}$ and $\simeq{\pm}18$\,cm$^{-1}$ in the high-temperature $\Gamma_{4}$ phase, but only one line at $\simeq{\pm}7$\,cm$^{-1}$ in the low-temperature $\Gamma_{2}$ magnetic structure.
The highest Raman intensities for the $\simeq{\pm}7$\,cm$^{-1}$ mode are observed in the $a(bc)\overline{a}$ [$a(cb)\overline{a}$] for the $\Gamma_{4}$ and in the $c(ab)\overline{c}$ [$c(ba)\overline{c}$] polarization configurations for the $\Gamma_{2}$ magnetic structure.
\textcolor{newtext}{We note that the HeNe laser has a satellite line at about 7.1 cm$^{-1}$, which appears only in the Stokes shift and was accurately accounted for in our analysis of the experimental results.}
The excitation with a frequency $\simeq{\pm}18$\,cm$^{-1}$ clearly seen in the $\Gamma_{4}$ magnetic structure has the highest Raman intensity in the $c(ab)\overline{c}$ [$c(ba)\overline{c}$] polarization configuration, while we could not reliably detect it in the $\Gamma_{2}$ phase.
Therefore, the selection rules from the Raman tensors~\eqref{eq:raman_tensors_magnon_Gamma2} and~\eqref{eq:raman_tensors_magnon_Gamma4} and typical values of frequencies allow us to assign these excitations to qFM ($\simeq{\pm}7$\,cm$^{-1}$) and qAFM ($\simeq{\pm}18$\,cm$^{-1}$) magnon modes.


\subsection{Spin-reorientation transition}

\begin{figure}
\centering
\includegraphics[width=1\columnwidth]{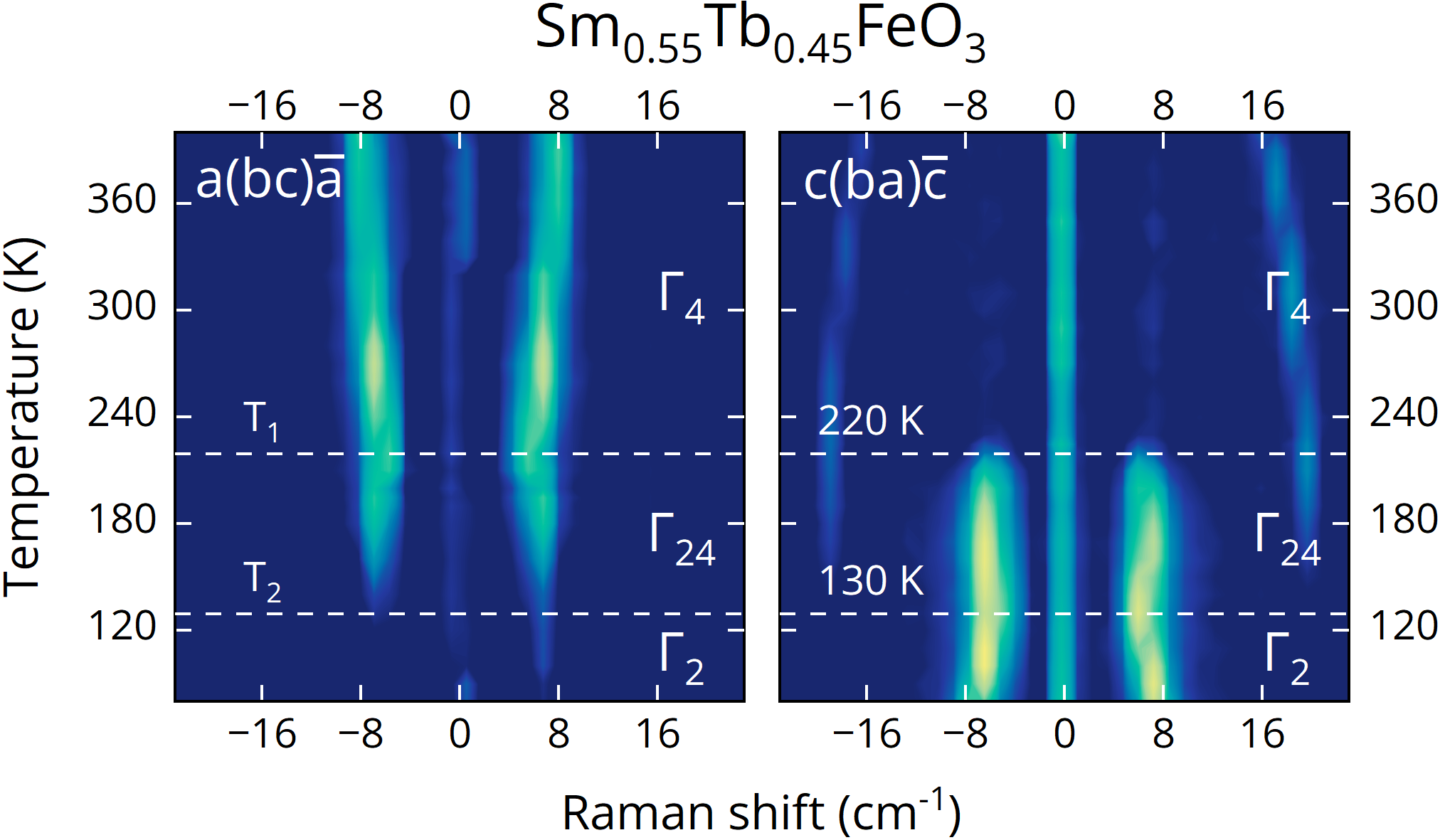}
\caption{\label{fig:magnon_map}
The temperature color map of the polarized Raman spectra of quasi-ferromagnetic (qFM, $\simeq{\pm7}$\,cm$^{-1}$) and quasi-antiferromagnetic (qAFM, $\simeq{\pm}18$\,cm$^{-1}$) modes in \STFO{} in the $a(bc)\overline{a}$ (left panel) and $c(ab)\overline{c}$ (right panel) polarization configurations which are given in Porto's notation as described in the text in the $Pbnm$ setting.
The $\Gamma_{4}$, $\Gamma_{24}$ and $\Gamma_{2}$ magnetic structures are separated by the white dashed lines.
The intensity is given on a log scale.
}
\end{figure}

Next, we measured the temperature dependence of the Raman spectra from 400 to 78\,K in the $a(cb)\overline{a}$ and $c(ba)\overline{c}$ polarization configurations. 
It is clearly seen that in the $a(bc)\overline{a}$ polarization configuration (left panel in Fig.~\ref{fig:magnon_map}) the qFM magnon mode is active at temperatures above $\simeq130$\,K, whereas above and below $\simeq130$\,K we could not observe any evidence of the qAFM magnon mode.
On the contrary, in the $c(ba)\overline{c}$ polarization configuration (right panel in Fig.~\ref{fig:magnon_map}) the qAFM magnon mode has been detected at temperatures above $\simeq130$\,K, and below $\simeq220$\,K the qFM mode becomes active. 
This allows us to conclude that the spin-reorientation transitions $\Gamma_{4}(G_{a}F_{c}) \xleftrightarrow{T_{1}} \Gamma_{24}(G_{ac}F_{ac}) \xleftrightarrow{T_{2}} \Gamma_{2}(G_{c}F_{a})$ with $T_{1} \simeq 220$\,K and $T_{2} \simeq 130$\,K are realized in \STFO.

We note that according to the literature, the Raman scattering from the qAFM magnon mode has not been detected in the $\Gamma_{2}$ magnetic structure in other orthoferrites, e.g. in $\mathrm{ErFeO}_{3}$~\cite{koshizuka1980inelastic,white1982light,koshizuka1988raman}, $\mathrm{SmFeO}_{3}$~\cite{koshizuka1988raman}, $\mathrm{TmFeO}_{3}$ and $\mathrm{TbFeO}_{3}$~\cite{venugopalan1983raman,venugopalan1985magnetic}.
However, the qAFM magnon mode is clearly seen in the $\Gamma_{2}$ magnetic structure in the magnetodipole absorption~\cite{stanislavchuk2017far}.
On the other hand, for the $\Gamma_{1}$ magnetic structure, observed at low temperatures in $\mathrm{DyFeO}_{3}$, the qAFM magnon mode has been revealed in the Raman scattering~\cite{koshizuka1988raman} and the THz experiments~\cite{stanislavchuk2016magnon,balbashov1985high}.  

\begin{figure}
\centering
\includegraphics[width=1\columnwidth]{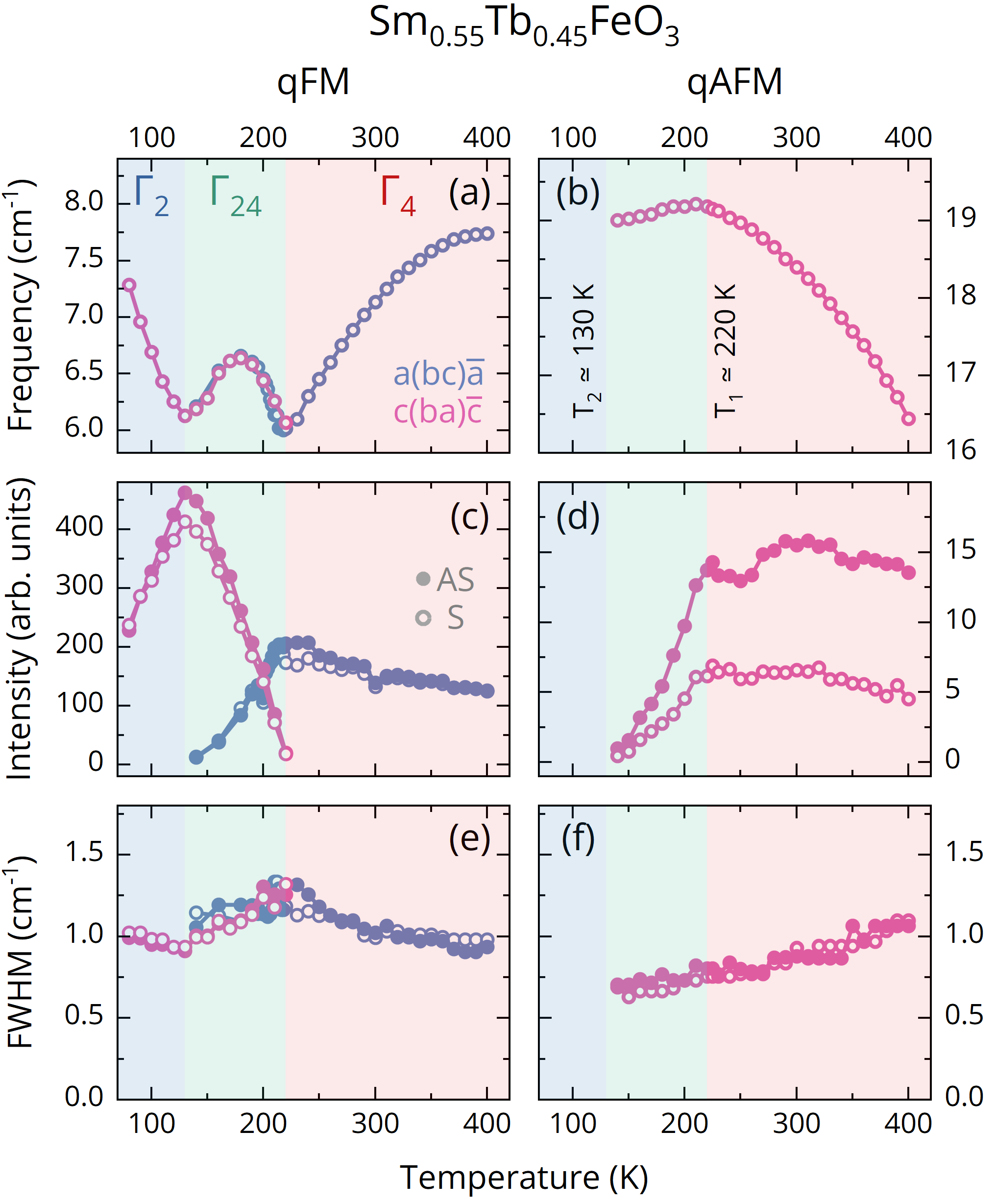}
\caption{\label{fig:magnon_tdep}
The temperature dependences of (a), (b)~frequency, (c), (d)~intensity and (e), (f)~full widths at half-maximum (FWHM) of the anti-Stokes (AS, closed circles) and Stokes (S, open circles) Raman lines in the $a(bc)\overline{a}$ and $c(ba)\overline{c}$ polarization configurations of the qFM and qAFM magnon excitations, respectively.
The blue, green, and red shaded areas correspond to the $\Gamma_{2}$ ($T < 130$\,K), $\Gamma_{24}$ ($130\mathrm{\,K} < T < 220\mathrm{\,K}$), and $\Gamma_{4}$ ($T > 220$\,K) magnetic configurations, respectively.
Intensities were not Bose corrected.
}
\end{figure}







To better understand the temperature behavior of magnetic excitations across spin-reorientation transitions, we fit the Raman spectra from Fig.~\ref{fig:magnon_map} using the procedure described earlier.
The results are shown in Figs.~\ref{fig:magnon_tdep}(a)--\ref{fig:magnon_tdep}(g).
The frequency of the qFM magnon mode shows a soft mode behavior and has two minima at temperatures $T_{1}\simeq220$\,K and $T_{2}\simeq130$\,K as can be seen in Fig.~\ref{fig:magnon_tdep}(a).
Moreover, the $\omega_{\mathrm{qFM}}$ at $T_{1}$ and $T_{2}$ has non-zero values, which is related to a magnetoelastic gap in the magnon spectrum ($\simeq 6$\,cm$^{-1}$) due to the magnetoelastic interaction of the quasi-ferromagnetic mode with acoustic phonon~\cite{turov1983broken}.
This soft mode behavior of the qFM magnon mode frequency was previously observed in Raman scattering on other orthoferrites with the $\Gamma_{4}(G_{a}F_{c}) \xleftrightarrow{T_{1}} \Gamma_{24}(G_{ac}F_{ac}) \xleftrightarrow{T_{2}} \Gamma_{2}(G_{c}F_{a})$ transition, e.g. in $\mathrm{ErFeO}_{3}$~\cite{koshizuka1980inelastic}, $\mathrm{TmFeO}_{3}$~\cite{venugopalan1983raman,venugopalan1985magnetic}.
At that, the qAFM mode is not a soft mode and its frequency increases at cooling in the $\Gamma_{4}$ magnetic structure and remains almost constant after the transition to the intermediate $\Gamma_{24}$ phase, as shown for the $c(ba)\overline{c}$ polarization configuration in Fig.~\ref{fig:magnon_tdep}(b).

\begin{table}
\caption{\label{tab:Gamma24} Temperatures $T_{1}$ and $T_{2}$ (in K) of the spin-reorientation transitions $\Gamma_{4} \xleftrightarrow{T_{1}} \Gamma_{24} \xleftrightarrow{T_{2}} \Gamma_{2}$ and temperature ranges $\Delta{T} = T_{1} - T_{2}$ in which the intermediate magnetic structure $\Gamma_{24}$ is realized for the earth-rare orthoferrites.}
\begin{ruledtabular}
\begin{tabular}{cccc}
   & $T_{1}$ & $T_{2}$ & $\Delta{T} = T_{1} - T_{2}$ \\ \midrule
$\mathrm{TbFeO}_{3}$\footnote{Refs.~\cite{nikolov1994mossbauer,balbashov1995submillimeter,artyukhin2012solitonic,stanislavchuk2017far}.}  & 9   & 8   & 1 \\
$\mathrm{TmFeO}_{3}$\footnote{Ref.~\cite{venugopalan1985magnetic}.}                                                                       & 94  & 84  & 10 \\
$\mathrm{ErFeO}_{3}$\footnote{Ref.~\cite{koshizuka1988raman}.}                                                                            & 103 & 90  & 13 \\
$\mathrm{Sm}_{0.7}\mathrm{Er}_{0.3}\mathrm{FeO}_{3}$\footnote{Ref.~\cite{fitzky2021ultrafast}.}                                           & 330 & 310 & 20 \\
$\mathrm{NdFeO}_{3}$\footnote{Ref.~\cite{li2019spin}.}                                                                                    & 170 & 107 & 63 \\
$\mathrm{PrFeO}_{3}$$^\text{e}$                                                                                                           & 10  & 6.5 & 3.5 \\
$\mathrm{Sm}_{0.4}\mathrm{Er}_{0.6}\mathrm{FeO}_{3}$\footnote{Ref.~\cite{wu2018growth}.}                                                  & 210 & 175 & 35 \\
$\mathrm{Sm}_{0.7}\mathrm{Tb}_{0.3}\mathrm{FeO}_{3}$\footnote{Ref.~\cite{wang2015single}.}                                                & 330 & 310 & 20 \\
$\mathrm{SmFeO}_{3}$\footnote{Ref.~\cite{weber2022emerging}.}                                                                             & 360 & 330 & 30 \\
\STFO{}\footnote{This work.}                                                                                                              & 220 & 130 & 90 \\
$\mathrm{Sm}_{0.5}\mathrm{Tb}_{0.5}\mathrm{FeO}_{3}$\footnote{Ref.~\cite{zhao2016spin}.}                                                  & 250 & 150 & 100 
\end{tabular}
\end{ruledtabular}
\end{table}

It should be emphasized that the temperature range $\Delta{T} = T_{1} - T_{2} \simeq 90$\,K in which the intermediate magnetic structure $\Gamma_{24}$ is realized is unusually wide in \STFO{} compared to other rare-earth orthoferrites with $\Gamma_{4}(G_{a}F_{c}) \xleftrightarrow{T_{1}} \Gamma_{24}(G_{ac}F_{ac}) \xleftrightarrow{T_{2}} \Gamma_{2}(G_{c}F_{a})$ transitions~\cite{li2025recent}, as can be seen for some crystals in Table~\ref{tab:Gamma24}.
It would be important to understand what causes such a wide temperature range $\Delta{T}$ in \STFO.
The frequencies of magnon modes in orthoferrites are described by the following expressions~\cite{vovk2025theory}: 
\begin{equation}
\label{eq:omega_qFM}
\omega_{\mathrm{qFM}} = \cfrac{\gamma}{m_{0}} \sqrt{H_{\mathrm{ex}} \, H_{ac}}, 
\quad \omega_{\mathrm{qAFM}} = \cfrac{\gamma}{m_{0}} \sqrt{H_{\mathrm{ex}} \, H_{ab}},
\end{equation}
where $\gamma$ is the gyromagnetic ratio, $m_{0}$ is the magnetization of the $\mathrm{Fe}^{3+}$ sublattices, and $H_{\mathrm{ex}}$ is the exchange field.
\textcolor{newtext}{According to Ref.~\cite{vovk2025theory}}, the effective anisotropy field $H_{ac}$ and $H_{ab}$ have the form
\begin{equation}
\label{eq:Hac}
    H_{ac}(T) = 
    \begin{cases}
        K_{ac}^{\mathrm{eff}}(T),                & T > T_{1}, \\
        2\, K_2\, \sqrt{- \cfrac{K^{\mathrm{eff}}_{ac}(T)}{K_{2}} \, \biggl[ 1 + \cfrac{K^{\mathrm{eff}}_{ac}(T)}{K_{2}} \biggr]}, & T_{2} < T < T_{1}, \\
        K_{ca}^{\mathrm{eff}}(T) = -K_{2} - K_{ac}^{\mathrm{eff}}(T),      & T_{2} > T,
    \end{cases}
\end{equation}
and
\begin{equation}
    H_{ab}(T) = 
    \begin{cases}
        K^{0}_{ab}, & T > T_{1}, \\
        K^{0}_{ab} + \biggl[1 + \cfrac{K^{\mathrm{eff}}_{ac}(T)}{K_{2}} \biggr] \, (K_{2}'' + K_{2}), & T_{2} < T < T_{1}, \\
        K^{0}_{ab} + K_{2}'' + K_{2}, & T_{2} > T,
    \end{cases}
\end{equation}
which includes the anisotropy function
\begin{equation}
\label{eq:K_ac}
\begin{gathered}
K_{ac}^{\mathrm{eff}}(T) = K_{ac}^{0} - \frac{(\Delta_{\mathrm{ex}}^{0})^{2}}{(\tilde{T} - \tilde{\lambda}_{\mathrm{f}})^{2}}\biggl[\tilde{\lambda}_{\mathrm{f}} - \frac{2}{(\tilde{T} - \tilde{\lambda}_{\mathrm{f}})^{2}}\biggl], \\
K_{ca}^{\mathrm{eff}}(T) = - 2 \, K_2 - K_{ac}^{0} + \frac{(\Delta_{\mathrm{ex}}^{0})^{2}}{(\tilde{T} - \tilde{\lambda}_{\mathrm{f}})^{2}}\biggl[\tilde{\lambda}_{\mathrm{f}} + \frac{2}{(\tilde{T} - \tilde{\lambda}_{\mathrm{f}})^{2}}\biggl],
\end{gathered} 
\end{equation}
and crystallographic anisotropy constants $K^{0}_{ac}$, $K^{0}_{ab}$, $K_{2}$, $K_{2}''$.
Here, $\Delta^{0}_{\mathrm{ex}}$ is the exchange parameter,
$\Delta_{\mathrm{cf}}$ is the crystal \textcolor{newtext}{field} splitting,
$\Delta_{\mathrm{R}}(T) = \sqrt{\Delta^{2}_{\mathrm{ex}}(T) + \Delta^{2}_{\mathrm{cf}}}$ is the energy splitting parameter,
$\tilde{T} = \Delta_{\mathrm{R}} \, / \, \tanh{(\Delta_{\mathrm{R}} \, / T)}$ is the reduced temperature and 
$\tilde{\lambda}_{\mathrm{f}}$ is the parameter, which represents $f$--$f$ exchange interaction.  

The temperature dependence of the magnon mode frequencies $\omega_{\mathrm{qFM}}(T)$ and $\omega_{\mathrm{qAFM}}(T)$ in \STFO{} can be fairly described using the parameters $K_{ac}^{0} = 0.326$\,K, $K_{2}=0.128$\,K, $\Delta_{\mathrm{cf}} = 67$\,K, and $\Delta_{\mathrm{ex}}^{0} = 5.55$\,K.
These parameters lie within the range predicted by theory, as detailed in Ref.~\cite{zvezdin1985rare}, and are in good agreement with the Refs.~\cite{vovk2025theory,buchelnikov1996magnetoacoustics,balbashov1989anomalies}, where similar parameters for other orthoferrites have been reported.
Since here we deal with the mixed rare-earth orthoferrite, these parameters are understood as ``effective'' and mostly refer to the $\mathrm{Sm}^{3+}$ ions, for which $\Delta_{\mathrm{cf}}$ is much larger than for $\mathrm{Tb}^{3+}$.

To understand further the origin of the spin-reorientation transition temperature range (i.e. interval between $T_{1}$ and $T_{2}$), let us assume a high-temperature approximation valid for $T_{1,2} \gg 10$\,K, so that the parameters responsible for the low-temperature $f$--$f$ exchange can be neglected.
The spin-reorientation transition is caused by the zero crossing of the effective anisotropy functions at temperatures $T_{1}$ and $T_{2}$.
Then Eq.~\eqref{eq:K_ac} for the effective anisotropies at the phase transition temperatures from Ref.~\cite{vovk2025theory} will take the following form:
\begin{equation}
\label{eq:Kac_Kca}
\begin{gathered}
    K_{ac}^{\mathrm{eff}}(T_{1}) = K_{ac}^{0} - 2 \, (\Delta_{\mathrm{ex}}^{0})^{2} \, \frac{\tanh{(\Delta_{\mathrm{cf}}/T_{1}})}{\Delta_{\mathrm{cf}}} = 0, \\
    K_{ca}^{\mathrm{eff}}(T_{2}) = -2 \, K_{2} - K_{ac}^{0} + 2 \, (\Delta_{\mathrm{ex}}^{0})^{2} \, \frac{\tanh{(\Delta_{\mathrm{cf}}/T_{1}})}{\Delta_{\mathrm{cf}}} = 0
\end{gathered}
\end{equation}
By solving these equations [Eq.~\eqref{eq:Kac_Kca}] with respect to $T_{1}$ and $T_{2}$, one obtains the following solutions:
\begin{equation}
\label{eq:DeltaT_Gamma24}
\begin{gathered}
    T_{1} = \frac{\Delta_{\mathrm{cf}}}{\arctanh{\Bigr[\cfrac{K^{0}_{ac} \, \Delta_{\mathrm{cf}}}{2 \, (\Delta^{0}_{\mathrm{ex}})^{2}}}\Bigr]}, \\
    T_{2} = \frac{\Delta_{\mathrm{cf}}}{\arctanh{\Bigr[\cfrac{(2 \, K_{2} + K^{0}_{ac}) \, \Delta_{\mathrm{cf}}}{2 \, (\Delta^{0}_{\mathrm{ex}})^{2}}}\Bigr]}.
\end{gathered}
\end{equation} 

It turns out to be convenient to analyze the ratio between $T_{1}$ and $T_{2}$, which has the following form:
\begin{equation}
\label{eq:T1T2ratio}
\frac{T_{1}}{T_{2}} = \frac{\arctanh{\Bigr[\cfrac{K^{0}_{ac} \, \Delta_{\mathrm{cf}}}{2 \, (\Delta^{0}_{\mathrm{ex}})^{2}}}\Bigr]}{\arctanh{\Bigr[\cfrac{(2 \, K_{2} + K^{0}_{ac}) \, \Delta_{\mathrm{cf}}}{2 \, (\Delta^{0}_{\mathrm{ex}})^{2}}}\Bigr]}.
\end{equation}
From this ratio, it can be seen that by setting cubic anisotropy $K_{2}$ to zero, the ratio will be equal to 1, corresponding to the scenario with only one second-order phase transition occurring at $T_{1}$. 
In the course of $\Gamma_{4} \xleftrightarrow{T_{1}} \Gamma_{24} \xleftrightarrow{T_{2}} \Gamma_{2}$ spin-reorientation transitions, which manifest themselves as two second-order phase transitions, at $T_{1}$ and $T_{2}$ and is observed in most of the orthoferrites, the value of $K_{2}$ must be positive~\cite{belov1976spin}.
For our parameters, the value of this ratio is $\Delta{T} \simeq 0.4967$.
By varying parameters included in Eq.~\eqref{eq:T1T2ratio} individually by +10\% while keeping the others constant, we obtain that for $K_{ac}^{0} = 0.358$\,K, the ratio changes to $\Delta{T} \simeq 0.509$; for $K_{2} = 0.1408$\,K it becomes $\Delta{T} \simeq 0.467$; for $\Delta_{\mathrm{cf}} = 73.6$\,K we evaluated a value $\Delta{T} \simeq 0.479$; and for $\Delta_{\mathrm{ex}}^{0} = 6.1$\,K it has taken on a value $\Delta{T} \simeq 0.519$.
Thus, the change in the anisotropy constant $K_{2}$ has a stronger effect on the spin-reorientation transition temperature range $\Delta{T}$ than corresponding changes in $K_{ac}^{0}$, $\Delta_{\mathrm{cf}}$ and $\Delta_{\mathrm{ex}}^{0}$, even though they are all of the same order.
Therefore, it can be concluded that the existence of the spin-reorientation transition is primarily defined by the cubic anisotropy $K_{2}$.
The temperature interval is mostly determined by the balancing of values of the cubic anisotropy $K_{2}$ and crystallographic anisotropy $K_{ac}^{0}$ as well as their ratio to the crystal field $\Delta_{\mathrm{cf}}$ and exchange interaction parameter $\Delta_{\mathrm{ex}}^{0}$.
The dilution of \SFO{} with $\mathrm{Tb}^{3+}$ ions reduces the concentration of $\mathrm{Sm}$ ions and introduces local defects into the magnetic texture, which in turn modifies the values of effective anisotropy parameters.

\begin{figure}
\centering
\includegraphics[width=1\columnwidth]{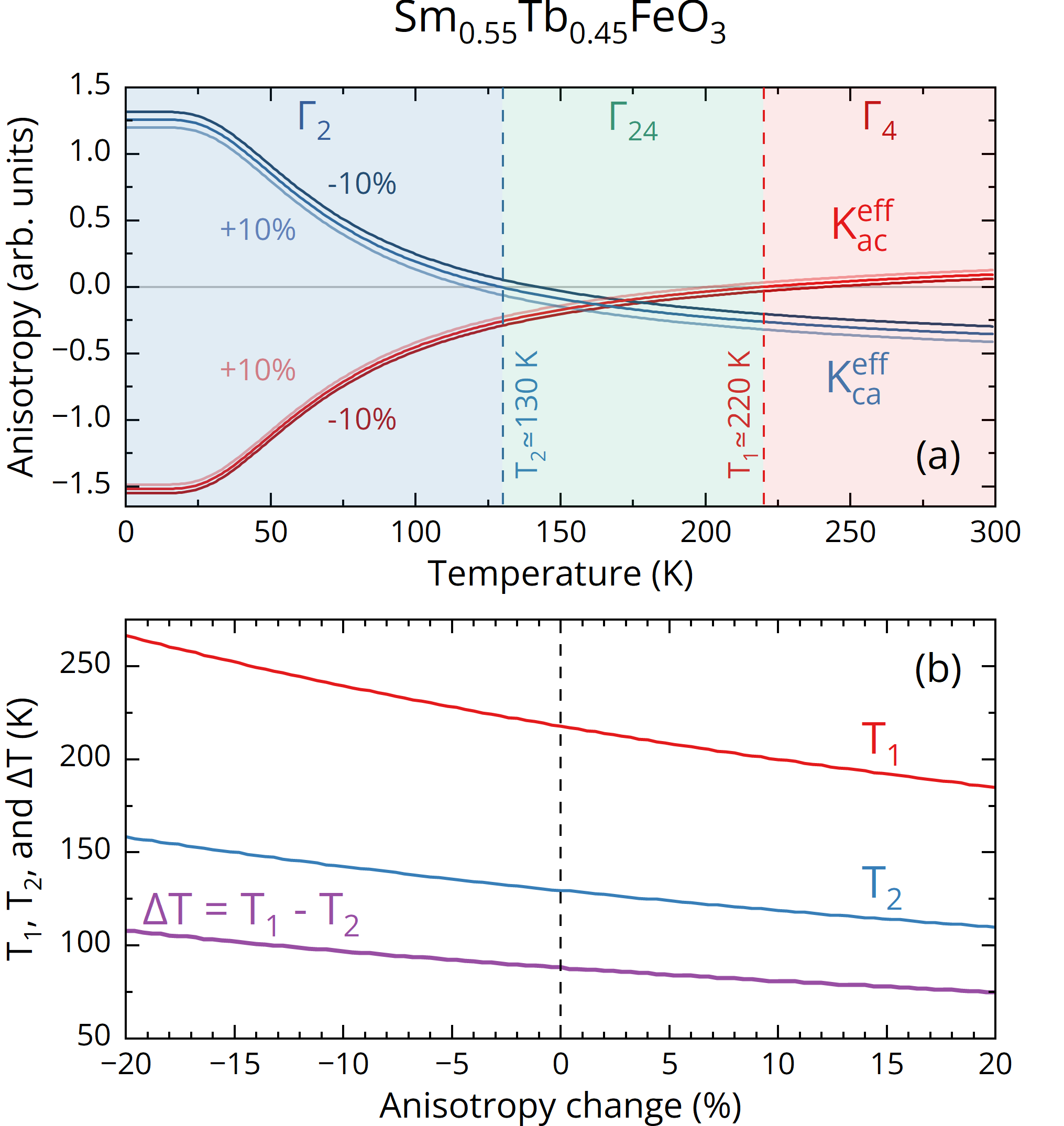}
\caption{\label{fig:T1T2}
(a)~The temperature dependences over the spin-reorientation transitions interval for the effective anisotropy functions $K_{ca}^{\mathrm{eff}}(T)$ and $K_{ac}^{\mathrm{eff}}(T)$, which determine the temperatures of the transitions $\Gamma_{4} \xleftrightarrow{T_{1}} \Gamma_{24} \xleftrightarrow{T_{2}} \Gamma_{2}$.
The change by $\pm10\%$ in the anisotropies leads to a change in $T_{1} \simeq 193-232$\,K and $T_{2} \simeq 121-145$\,K. 
(b)~Temperatures of spin-reorientation transitions $T_{1}$ and $T_{2}$ and temperature ranges $\Delta{T} = T_{1} - T_{2}$ of the intermediate magnetic structure $\Gamma_{24}$ as functions of the magnetic anisotropy.
}
\end{figure}

Figure~\ref{fig:T1T2}(a) shows the temperature dependences of the effective anisotropies $K_{ca}^{\mathrm{eff}}(T)$ and $K_{ac}^{\mathrm{eff}}(T)$ according to Eq.~\eqref{eq:K_ac} when only anisotropy parameters $K_{2}$ and $K_{ac}^{0}$ are varied by $\pm10\%$.
The crossing of the zero by $K_{ac}^{\mathrm{eff}}(T)$ and $K_{ca}^{\mathrm{eff}}(T)$ occurs at the transition temperatures $T_{1}$ and $T_{2}$.
It can be seen that a decrease in anisotropies leads to a lowering in these temperatures. 
Moreover, the temperature $T_{1}$ changes with a change in anisotropy more strongly than $T_{2}$, which in turn leads to an increase in the temperature range $\Delta{T}$, as shown in Fig.~\ref{fig:T1T2}(b).
Thus, in the mixed sample, the value of the effective anisotropy is different from the pure orthoferrites due to the random distribution of the $\mathrm{Sm}^{3+}$ and $\mathrm{Tb}^{3+}$ ions, making the perfect magnetic structure defected, which may explain the broad temperature interval of the spin-reorientation.

The intensities for the anti-Stokes and Stokes lines of the qFM mode have close temperature behavior as can be seen in Fig.~\ref{fig:magnon_tdep}(c).
They depend weakly on the temperature for $a(bc)\overline{a}$ polarization configuration in the $\Gamma_{4}$ magnetic structure, and at the transition to the intermediate $\Gamma_{24}$ magnetic structure, the intensity begins to decrease sharply to extremely low values at cooling.
In the $c(ab)\overline{c}$ polarization configuration the intensity of the qFM mode begins to rapidly increase at cooling in the $\Gamma_{24}$ phase, but at the transition in the $\Gamma_{2}$ magnetic structure, the trend changes and it gradually decreases.   
At that, the FWHM of the qFM mode is about 1.0\,cm$^{-1}$ and weakly dependent on the temperature for both polarization configurations [see Fig.~\ref{fig:magnon_tdep}(e)]. 

As already stated, the qAFM mode does not appear in any polarization configurations in the $\Gamma_{2}$ magnetic structure.
It is clearly seen in Fig.~\ref{fig:magnon}, the intensity of the Stokes line is greater than the anti-Stokes one of the qAFM mode in the  $c(ba)\overline{c}$ polarization configuration in the $\Gamma_{4}$ structure.
These intensities are essentially unchanged in the $\Gamma_{4}$ structure, whereas in the $\Gamma_{24}$ phase at cooling they sharply decrease to almost very low values as shown in Fig.~\ref{fig:magnon_tdep}(d).
The FWHM decreases at cooling and does not change significantly at $\Gamma_{4}(G_{a},F_{c}) \xleftrightarrow{T_{1}} \Gamma_{24}(G_{ac},F_{ac})$ magnetic transition after which it rises, which is probably due to the difficulty of determining the FWHM of lines with low intensities. 

\begin{figure}
\centering
\includegraphics[width=1\columnwidth]{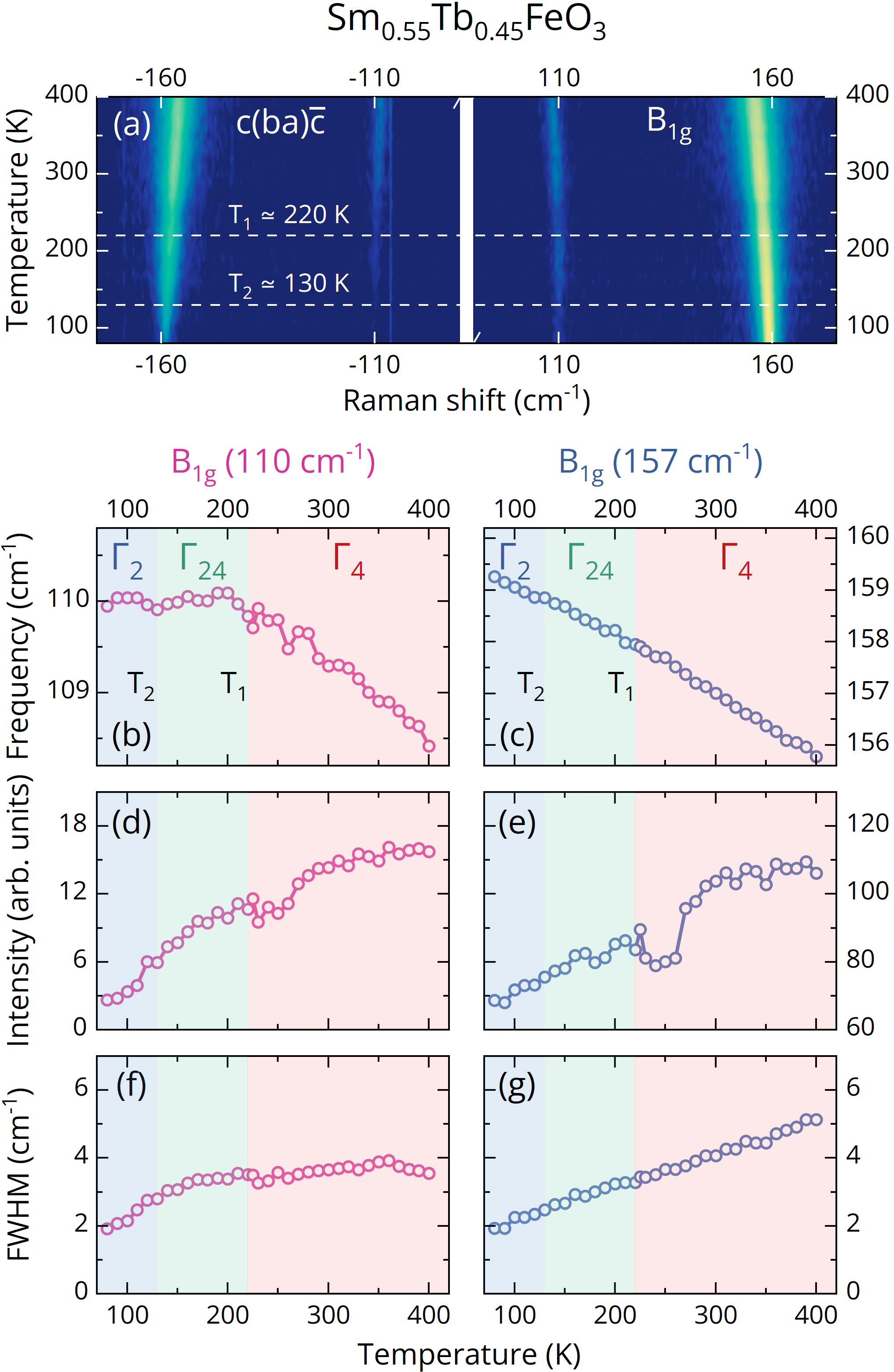}
\caption{\label{fig:phonon_tdep}
(a)~The temperature color map of the polarized Raman spectra for the $c(ab)\overline{c}$ polarization configuration in the range that involves low-frequency $B_{1g}$ phonons.
The intensity is given on a log scale.
The temperature dependences of (b), (c)~frequency, (d), (e)~intensity and (f), (g)~full widths at half maximum (FWHM) of the Stokes lines of the two low-frequency $B_{1g}$ phonon from panel (a).
The blue, green, and red shaded areas correspond to the $\Gamma_{2}$ ($T < 130$\,K), $\Gamma_{24}$ ($130\mathrm{\,K} < T < 220\mathrm{\,K}$), and $\Gamma_{4}$ ($T > 220$\,K) magnetic configurations, respectively.
Intensities were not Bose corrected.
}
\end{figure}

The temperature dependences of the Raman scattering of magnetic excitations were measured in the range from -180 to 180\,cm$^{-1}$, and they allowed us to trace the behavior of two $B_{1g}$ phonons ($\simeq 110\,\mathrm{cm}^{-1}$ and $\simeq 157\,\mathrm{cm}^{-1}$) which are active in the $c(ab)\overline{c}$ polarization configuration [see Figs.~\ref{fig:raman}(d), \ref{fig:phonon_tdep}(a), and Table~\ref{tab:raman_phonons}].  
The Stokes lines in the Raman spectra of these phonons were fitted using the procedure described above, which allowed us to derive the temperature dependences of the frequency, intensity, and FWHM as shown in Fig.~\ref{fig:phonon_tdep}(b)--~\ref{fig:phonon_tdep}(g), respectively.
All of them demonstrate a typical behavior for Raman-active phonons, with the frequency increasing and the intensity and FWHM decreasing.
We suppose that the drop in the intensity of the phonon in Fig.~\ref{fig:phonon_tdep}(e) at around 250\,K is likely an experimental artifact.
It is clearly seen that the parameters of these $B_{1g}$ Raman-active phonons do not show any noticeable change at $T_{1} \simeq 220$\,K and $T_{2} \simeq 130$\,K thus allowing us to conclude that this phonon is not susceptible to the $\Gamma_{4}(G_{a},F_{c}) \xleftrightarrow{T_{1}} \Gamma_{24}(G_{ac},F_{ac}) \xleftrightarrow{T_{2}} \Gamma_{2}(G_{c},F_{a})$ spin-reorientation transitions.
This conclusion is further supported by the fact that the involvement of $\mathrm{Fe}^{3+}$ ions in Raman-active phonons is forbidden by symmetry~\cite{dubrovin2024lattice}.
The absence of any evidence of spin-phonon coupling for Raman-active phonons at the antiferromagnetic ordering of $\mathrm{Fe}^{3+}$ ions below $T_{N}$ has been observed previously in rare-earth orthoferrites~\cite{weber2022emerging}.
Moreover, the lattice parameters of orthoferrites do not any clear change at $T_{N}$~\cite{pavlovska2016thermal}.
However, it was reported in Ref.~\cite{panchwanee2019temperature} that slight changes in the frequency of some Raman-active phonons at the spin-reorientation transitions were detected in polycrystalline \SFO{} samples.

It is known that the heavy rare-earth ions are significantly involved in the vibration of the lowest-frequency Raman-active phonon in orthoferrites~\cite{dubrovin2024lattice}.
The magnetism of $\mathrm{Fe}^{3+}$ iron ions induces a non-spontaneous polarization of the rare-earth ion spins, which is manifested in the shift of phonon frequencies and the appearance of new phonon lines as a result of possible symmetry lowering~\cite{li2023terahertz,weber2022emerging}. 
A further important result of our experimental study is the absence of any evidence of new $B_{1g}$ phonon lines in the temperature color map in Fig.~\ref{fig:phonon_tdep}(a) and no apparent deviation of parameters of the lowest-frequency $B_{1g}$ phonons from anharmonic behavior, as can be seen in Figs.~\ref{fig:phonon_tdep}(b)--\ref{fig:phonon_tdep}(g).
This strongly indicates that the rare-earth ions in \STFO{} do not experience non-spontaneous polarization under the action of the magnetic iron sublattice at the specified temperature range. 

\section{Concluding remarks}
In summary, we have scrutinized the lattice and magnetic dynamics at the center of the Brillouin zone of the rare-earth orthoferrite \STFO{} single crystal by the polarized spectroscopic infrared and Raman scattering technique. 
The obtained experimental results on the lattice dynamics were supported by the first-principles calculations, which allowed us to reliably identify the parameters of the most infrared- and Raman-active phonons.
Further, we explored the Raman spectra of the quasi-ferromagnetic mode and quasi-antiferromagnetic mode in all main polarization configurations at 78\,K and 293\,K which correspond to the $\Gamma_{2}$ and $\Gamma_{4}$ magnetic structures in this compound, respectively.

We studied the temperature dependence from 78\,K to 400\,K of the spectra in which the Raman scattering of the quasi-ferromagnetic and quasi-antiferromagnetic modes is most intense in the  $\Gamma_{2}$ and $\Gamma_{4}$ magnetic structures.
As a result, it was found that a sequence of spin-reorientation transitions $\Gamma_{4} \xleftrightarrow{T_{1}} \Gamma_{24} \xleftrightarrow{T_{2}} \Gamma_{2}$ is observed at the temperatures $T_{1} \simeq 220$\,K and $T_{2} \simeq 130$\,K in the \STFO.
At that, the quasi-ferromagnetic mode is a soft mode the frequency of which does not go to zero due to magnetoelastic interaction with acoustic phonons.
The theoretical analysis demonstrates that the temperature range of the intermediate magnetic phase $\Gamma_{24}$ is primarily controlled by the magnetocrystalline anisotropy. 
The random distribution of rare-earth $\mathrm{Sm}^{3+}$ and $\mathrm{Tb}^{3+}$ ions results in formation of defects into the magnetic lattice of \STFO{} compared to pure orthoferrites.
These defects lead to a change in the magnetocrystalline anisotropy, resulting in an increased temperature range $\Delta{T} = T_{1} - T_{2} \simeq 90$\,K for the intermediate $\Gamma_{24}$ phase.
We have not found any clear evidence for Raman scattering of the quasi-antiferromagnetic mode in the $\Gamma_{2}$ magnetic structure. 
Furthermore, we have revealed that the two lowest-frequency $B_{1g}$ Raman-active phonons do not show any obvious changes at the both spin-reorientation transitions, which we attribute to the fact that the involvement of the $\mathrm{Fe}$ ions in the Raman-active phonons is forbidden by symmetry.  
We believe that our findings will stimulate further experimental research into the booming field of nonlinear phononic, magnetophononic and spin-related effects in the other rare-earth orthoferrites~\cite{gareev2024optical,leenders2024canted,zhang2024spin,huang2024extreme,zhang2024upconversion,zhang2024coupling,afanasiev2021ultrafast,nova2017effective}.

\begin{table*}
\caption{\label{tab:dft_phonons} Calculated frequencies (cm$^{-1}$) of the phonons at the Brillouin zone center in rare-earth orthoferrites \TFO{}, \SFO{}, and \STFO{}.
The procedure for estimating the phonon frequencies for \STFO{} using those for \TFO{} and \SFO{} is described in detail in the main text.
}

\begin{ruledtabular}
\begin{tabular}{ccccccccc>{\hspace{10pt}}c>{\hspace{10pt}}c}
                            \multirow{2}{*}{Sym.} 
                            & \multirow{2}{*}{\TFO} & \multirow{2}{*}{\SFO} 
                            & \multirow{2}{*}{\STFO} 
                            & \multirow{2}{*}{Sym.} 
                            &  \multicolumn{2}{c}{\TFO} 
                            & \multicolumn{2}{c}{\SFO} 
                            & \multicolumn{2}{c}{\STFO} 
                            \\
                            \cmidrule{6-7} \cmidrule{8-9} \cmidrule{10-11} 
                             & & & & & TO & LO & TO & LO & TO & LO \\
                            \cmidrule{1-4} \cmidrule{5-11}
\multirow{7}{*}{$A_{g}$}  & 111.8 & 110.3 & 111   & \multirow{7}{*}{$B_{1u}$} & 146.3 & 152   & 151.9 & 152.9 & 149.4 & 152.5 \\
                          & 131.8 & 134.2 & 133.1 &                           & 158.1 & 177.6 & 163.1 & 181.4 & 160.8 & 179.7 \\
                          & 253.1 & 240   & 245.9 &                           & 241.3 & 283   & 243   & 280   & 242.2 & 281.4 \\
                          & 330   & 319.3 & 324.1 &                           & 302.6 & 303.7 & 298.4 & 302.1 & 300.3 & 302.8 \\
                          & 402.4 & 378.6 & 389.3 &                           & 326.2 & 455.1 & 315.3 & 453.3 & 320.2 & 454.1 \\
                          & 408.8 & 402   & 405.1 &                           & 467.3 & 482.1 & 464.6 & 478.5 & 465.8 & 480.1 \\ 
                          & 487.6 & 471.7 & 478.9 &                           & 511   & 643.2 & 509.4 & 640.9 & 510.1 & 641.9 \\ \cmidrule{1-4} \cmidrule{5-11}
\multirow{7}{*}{$B_{1g}$} & 109.6 & 110.2 & 109.9 & \multirow{9}{*}{$B_{2u}$} & 104.4 & 104.7 & 107.8 & 108   & 106.3 & 106.5 \\
                          & 157.8 & 156.4 & 157   &                           & 185.9 & 186.7 & 184.6 & 184.8 & 185.2 & 185.7 \\
                          & 293.9 & 271.2 & 281.4 &                           & 224.5 & 256.3 & 223   & 253   & 223.7 & 254.5 \\
                          & 351.7 & 347.9 & 349.6 &                           & 288.8 & 295.6 & 277   & 293.7 & 282.3 & 294.6 \\
                          & 483.1 & 465.3 & 473.3 &                           & 297.7 & 336.7 & 295.3 & 329.5 & 296.4 & 332.7 \\
                          & 529.3 & 515.8 & 521.9 &                           & 336.7 & 401.4 & 329.5 & 397.3 & 332.7 & 399.1 \\
                          & 641.1 & 639.8 & 640.4 &                           & 417.2 & 494.5 & 409   & 480.8 & 412.7 & 487   \\ \cmidrule{1-4}
\multirow{5}{*}{$B_{2g}$} & 122.6 & 131.1 & 127.3 &                           & 506.6 & 527.3 & 495.9 & 517.5 & 500.7 & 521.9 \\
                          & 314.3 & 314.8 & 314.6 &                           & 528.5 & 642.7 & 517.9 & 632.3 & 522.7 & 637   \\ \cmidrule{5-11}
                          & 412.4 & 407.8 & 409.9 & \multirow{9}{*}{$B_{3u}$} & 114.7 & 116.1 & 115.1 & 116.4 & 114.9 & 116.3 \\ 
                          & 463.7 & 453.4 & 458   &                           & 171.5 & 176   & 171.8 & 176   & 171.7 & 176   \\ 
                          & 667.8 & 670.2 & 669.1 &                           & 221.8 & 275.9 & 220.1 & 273.8 & 220.9 & 274.7 \\ \cmidrule{1-4}
\multirow{5}{*}{$B_{3g}$} & 140.9 & 152.9 & 147.5 &                           & 280.7 & 318.7 & 277.2 & 314.8 & 278.8 & 316.6 \\ 
                          & 239.8 & 227.4 & 233   &                           & 328.4 & 334   & 322.1 & 329.4 & 324.9 & 331.5 \\
                          & 347.8 & 342   & 344.6 &                           & 347.3 & 369.4 & 342.2 & 357.2 & 344.5 & 362.7 \\
                          & 409.6 & 402.3 & 405.6 &                           & 392.2 & 492   & 375.6 & 470.6 & 383.1 & 480.2 \\
                          & 622.8 & 621.1 & 621.9 &                           & 495.7 & 519.9 & 473   & 510.9 & 483.2 & 514.9 \\ \cmidrule{1-4}
                          &       &       &       &                           & 535.4 & 638.7 & 526.5 & 629.7 & 530.5 & 633.8 \\ \cmidrule{5-11}
\multirow{4}{*}{$A_{u}$}  & 79    & 84.5  & 82    & \multirow{4}{*}{$A_{u}$}  & 308   &       & 307.5 &       & 307.7 &       \\
                          & 155.6 & 158.2 & 157   &                           & 343.2 &       & 330.7 &       & 336.3 &       \\
                          & 182.9 & 182.5 & 182.7 &                           & 456.8 &       & 456.3 &       & 456.5 &       \\
                          & 220.5 & 225.4 & 223.2 &                           & 503.6 &       & 495.5 &       & 499.1 &       \\
\end{tabular}
\end{ruledtabular}
\end{table*}

\section*{Acknowledgements}
We thank N.\,A.\,Arkhipov for the help with the XRD orientation, Dr.~G.\,A.\,Gusev and Prof.~M.\,V.\,Zamoryanskaya for the XRF characterization of single crystals, and G.\,V.\,Osochenko for assistance in experimental research.
This work was supported by the Russian Science Foundation under Grant No. 24-72-00106, https://rscf.ru/en/project/24-72-00106/.
A.I.B. acknowledges the support of the Ministry of Science and Higher Education of the Russian Federation (Grant No.~FSWR-2024-0003).
I.A.E., A.N.S. and V.Yu.D. acknowledge the support of the Ministry of Science and Higher Education of the Russian Federation (Grant No.~FFUG-2024-0018).
A.M.K. acknowledges the support of the Ministry of Science and Higher Education of the Russian Federation (Grant No.~FFUG-2024-0037).
V.A.C. also acknowledges the support of the Ministry of Science and Higher Education of the Russian Federation (Grant No.~FEUZ-2023-0017).
N.N.N. acknowledges the support of the research project FFUU-2025-0004 of the Institute of Spectroscopy of the Russian Academy of Sciences.
A.W. and L.S. acknowledge the support of the National Natural Science Foundation of China (NSFC, Nos. 52272014, W2421113), and the International Partnership Program of Chinese Academy of Sciences (No. 030GJHZ2024093MI).
R.V.M. acknowledges the support of Royal Society International Exchanges 2021, grant IES$\backslash$R2$\backslash$212182.

\section*{Data availability}
The data that support the findings of this article are openly available~\cite{dataset}.

\bibliography{bibliography}

@article{venugopalan1985magnetic,
    title = {{M}agnetic and vibrational excitations in rare-earth orthoferrites: {A} {R}aman scattering study},
    author = {Venugopalan, S. and Dutta, Mitra and Ramdas, A. K. and Remeika, J. P.},
    journal = {Phys. Rev. B},
    volume = {31},
    issue = {3},
    pages = {1490--1497},
    numpages = {0},
    year = {1985},
    month = {Feb},
    publisher = {American Physical Society},
    doi = {10.1103/PhysRevB.31.1490},
    url = {https://link.aps.org/doi/10.1103/PhysRevB.31.1490}
}

@article{gupta2002lattice,
    title = {{L}attice dynamic investigation of {R}aman and infrared wavenumbers at the zone center of orthorhombic $\mathrm{RFeO}_{3}$ ({R} = {Tb}, {Dy}, {Ho}, {Er}, {Tm}) perovskites},
    author = {Gupta, H. C. and Kumar Singh, Manoj and Tiwari, L. M.},
    journal = {J. Raman Spectrosc.},
    volume = {33},
    number = {1},
    pages = {67--70},
    year = {2002},
    publisher = {Wiley Online Library},
    doi = {10.1002/jrs.805},
    url = {https://doi.org/10.1002/jrs.805}
}

@article{stanislavchuk2017far,
    title = {{F}ar-{IR} magnetospectroscopy of magnons and electromagnons in $\mathrm{TbFeO}_{3}$ single crystals at low temperatures},
    author = {Stanislavchuk, T. N. and Wang, Yazhong and Cheong, S.-W. and Sirenko, A. A.},
    journal = {Phys. Rev. B},
    volume = {95},
    issue = {5},
    pages = {054427},
    numpages = {11},
    year = {2017},
    month = {Feb},
    publisher = {American Physical Society},
    doi = {10.1103/PhysRevB.95.054427},
    url = {https://link.aps.org/doi/10.1103/PhysRevB.95.054427}
}

@article{weber2022emerging,
    title = {Emerging spin--phonon coupling through cross-talk of two magnetic sublattices},
    author = {Weber, Mads C. and Guennou, Mael and Evans, Donald M. and Toulouse, Constance and Simonov, Arkadiy and Kholina, Yevheniia and Ma, Xiaoxuan and Ren, Wei and Cao, Shixun and Carpenter, Michael A. and Dkhil, Brahim and Fiebig, Manfred and Kreisel, Jens},
    journal = {Nature Commun.},
    volume = {13},
    number = {1},
    pages = {1--7},
    year = {2022},
    publisher = {Nature Publishing Group},
    doi = {10.1038/s41467-021-27267-8},
    url = {https://doi.org/10.1038/s41467-021-27267-8}
}

@article{weber2016raman,
    title = {Raman spectroscopy of rare-earth orthoferrites $\mathit{R}\mathrm{FeO}_{3}$ ($\it{R}$ = $\mathrm{La}$, $\mathrm{Sm}$, $\mathrm{Eu}$, $\mathrm{Gd}$, $\mathrm{Tb}$, $\mathrm{Dy}$)},
    author = {Weber, Mads Christof and Guennou, Mael and Zhao, Hong Jian and \'I\~niguez, Jorge and Vilarinho, Rui and Almeida, Ab\'{\i}lio and Moreira, Joaquim Agostinho and Kreisel, Jens},
    journal = {Phys. Rev. B},
    volume = {94},
    issue = {21},
    pages = {214103},
    numpages = {8},
    year = {2016},
    month = {Dec},
    publisher = {American Physical Society},
    doi = {10.1103/PhysRevB.94.214103},
    url = {https://link.aps.org/doi/10.1103/PhysRevB.94.214103}
}

@article{artyukhin2012solitonic,
    title = {{S}olitonic lattice and {Y}ukawa forces in the rare-earth orthoferrite $\mathrm{TbFeO}_{3}$},
    author = {Artyukhin, Sergey and Mostovoy, Maxim and Jensen, Niels Paduraru and Le, Duc and Prokes, Karel and De Paula, Vin{\'\i}cius G and Bordallo, Heloisa N and Maljuk, Andrey and Landsgesell, Sven and Ryll, Hanjo and Klemke, Bastian and Paeckel, Sebastian and Kiefer, Klaus and Lefmann, Kim and Kuhn, Luise Theil and Argyriou, Dimitri N.},
    journal = {Nature Mater.},
    volume = {11},
    number = {8},
    pages = {694--699},
    year = {2012},
    publisher = {Nature Publishing Group},
    doi = {10.1038/nmat3358},
    url = {https://doi.org/10.1038/nmat3358}
}

@article{kroumova2003bilbao,
    title = {Bilbao crystallographic server: useful databases and tools for phase-transition studies},
    author = {Kroumova, E and Aroyo, M I and Perez-Mato, J M and Kirov, A and Capillas, C and Ivantchev, S and Wondratschek, H},
    journal = {Phase Transit.},
    volume = {76},
    number = {1-2},
    pages = {155--170},
    year = {2003},
    publisher = {Taylor \& Francis}
}

@article{ponosov2020lattice,
    title = {Lattice and spin excitations of $\mathrm{YFeO}_{3}$: {A} {R}aman and density functional theory study},
    author = {Ponosov, {\relax Yu} S and Novoselov, D. Y.},
    journal = {Phys. Rev. B},
    volume = {102},
    issue = {5},
    pages = {054418},
    numpages = {9},
    year = {2020},
    month = {Aug},
    publisher = {American Physical Society},
    doi = {10.1103/PhysRevB.102.054418},
    url = {https://link.aps.org/doi/10.1103/PhysRevB.102.054418}
}

@article{tokunaga2008magnetic,
    title = {{M}agnetic-{F}ield-{I}nduced {F}erroelectric {S}tate in $\mathrm{DyFeO}_{3}$},
    author = {Tokunaga, Y. and Iguchi, S. and Arima, T. and Tokura, Y.},
    journal = {Phys. Rev. Lett.},
    volume = {101},
    issue = {9},
    pages = {097205},
    numpages = {4},
    year = {2008},
    month = {Aug},
    publisher = {American Physical Society},
    doi = {10.1103/PhysRevLett.101.097205},
    url = {https://link.aps.org/doi/10.1103/PhysRevLett.101.097205}
}

@article{belov1976spin,
    title = {Spin-reorientation transitions in rare-earth magnets},
    author = {Belov, Konstantin P and Zvezdin, Anatolii K and Kadomtseva, Antonina M and Levitin, R Z},
    journal = {Sov. Phys. Uspekhi},
    volume = {19},
    number = {7},
    pages = {574},
    year = {1976},
    publisher = {IOP Publishing},
    doi = {10.1070/PU1976v019n07ABEH005274},
    url = {https://doi.org/10.1070/PU1976v019n07ABEH005274}
}

@article{nova2017effective,
  title = {An effective magnetic field from optically driven phonons},
  author = {Nova, Tobia F and Cartella, Andrea and Cantaluppi, Alice and F{\"o}rst, Michael and Bossini, Davide and Mikhaylovskiy, Rostislav V and Kimel, Aleksei V and Merlin, Roberto and Cavalleri, Andrea},
  journal = {Nature Phys.},
  volume = {13},
  number = {2},
  pages = {132--136},
  year = {2017},
  publisher = {Nature Publishing Group},
  doi = {10.1038/nphys3925},
  url = {https://doi.org/10.1038/nphys3925}
}

@article{afanasiev2021ultrafast,
    title = {Ultrafast control of magnetic interactions via light-driven phonons},
    author = {Afanasiev, Dmytro and Hortensius, J R and Ivanov, B A and Sasani, Alireza and Bousquet, Eric and Blanter, Y M and Mikhaylovskiy, R V and Kimel, A V and Caviglia, A D},
    journal = {Nature Mater.},
    volume = {20},
    number = {5},
    pages = {607--611},
    year = {2021},
    publisher = {Nature Publishing Group},
    doi = {10.1038/s41563-021-00922-7},
    url = {https://doi.org/10.1038/s41563-021-00922-7}
}

@article{kimel2005ultrafast,
    title = {Ultrafast non-thermal control of magnetization by instantaneous photomagnetic pulses},
    author = {Kimel, A V and Kirilyuk, A and Usachev, P A and Pisarev, R V and Balbashov, A M and Rasing, {\relax Th}},
    journal = {Nature},
    volume = {435},
    number = {7042},
    pages = {655--657},
    year = {2005},
    publisher = {Nature Publishing Group},
    doi = {10.1038/nature03564},
    url = {https://doi.org/10.1038/nature03564}
}

@article{kimel2009inertia,
    title = {Inertia-driven spin switching in antiferromagnets},
    author = {Kimel, A V and Ivanov, B A and Pisarev, R V and Usachev, P A and Kirilyuk, A and Rasing, {\relax Th}},
    journal = {Nature Phys.},
    volume = {5},
    number = {10},
    pages = {727--731},
    year = {2009},
    publisher = {Nature Publishing Group},
    doi = {10.1038/nphys1369},
    url = {https://doi.org/10.1038/nphys1369}
}

@article{wang2015single,
    title = {Single crystal growth and magnetic properties of $\mathrm{Sm}_{0.7}\mathrm{Tb}_{0.3}\mathrm{FeO}_{3}$ orthoferrite single crystal},
    author = {Wang, Bo and Zhao, Xiangyang and Wu, Anhua and Cao, Shixun and Xu, Jun and Kalashnikova, A. M. and Pisarev, R. V.},
    journal = {J. Magn. Magn. Mater.},
    volume = {379},
    pages = {192--195},
    year = {2015},
    publisher = {Elsevier},
    doi = {10.1016/j.jmmm.2014.12.030},
    url = {https://doi.org/10.1016/j.jmmm.2014.12.030}
}

@article{dovesi2014crystal14,
    title = {\textsc{CRYSTAL14}: {A} program for the \textit{ab initio} investigation of crystalline solids},
    author = {Dovesi, Roberto and Orlando, Roberto and Erba, Alessandro and Zicovich-Wilson, Claudio M and Civalleri, Bartolomeo and Casassa, Silvia and Maschio, Lorenzo and Ferrabone, Matteo and De La Pierre, Marco and d'Arco, Philippe and others},
    journal={Int. J. Quantum Chem.},
    volume = {114},
    number = {19},
    pages = {1287--1317},
    year = {2014},
    publisher = {Wiley Online Library},
    doi = {10.1002/qua.24658},
    url = {https://doi.org/10.1002/qua.24658}
}

@article{becke1993density,
    title = {Density-functional thermochemistry. {III}. {T}he role of exact exchange},
    author = {Becke, Axel D},
    journal = {J. Chem. Phys.},
    volume = {98},
    number = {7},
    pages = {5648--5652},
    year = {1993},
    publisher = {American Institute of Physics},
    doi = {10.1063/1.464913},
    url = {https://doi.org/10.1063/1.464913}
}

@article{white1982light,
    title = {Light scattering from magnetic excitations in orthoferrites},
    author = {White, R. M. and Nemanich, R. J. and Herring, Conyers},
    journal = {Phys. Rev. B},
    volume = {25},
    number = {3},
    pages = {1822},
    year = {1982},
    publisher = {American Physical Society},
    doi = {10.1103/PhysRevB.25.1822},
    url = {https://doi.org/10.1103/PhysRevB.25.1822}
}

@article{balbashov1985high,
    title = {High-frequency magnetic properties of dysprosium orthoferrite},
    author = {Balbashov, A. M. and Volkov, A. A. and Lebedev, S. P. and Mukhin, A. A. and Prokhorov, A. S.},
    journal = {Sov. Phys. JETP},
    volume = {61},
    pages = {573},
    year = {1985},
    url = {http://www.jetp.ras.ru/cgi-bin/dn/e_061_03_0573}
}

@article{li2023terahertz,
    title = {Terahertz spin dynamics in rare-earth orthoferrites},
    author = {Li, Xinwei and Kim, Dasom and Liu, Yincheng and Kono, Junichiro},
    journal = {Photonic. Ins.},
    volume = {1},
    number = {2},
    pages = {R05},
    year = {2023},
    publisher = {SPIE},
    doi = {10.3788/PI.2022.R05},
    url = {https://doi.org/10.3788/PI.2022.R05}
}

@article{coutinho2017structural,
    title = {Structural, vibrational and magnetic properties of the orthoferrites $\mathrm{LaFeO}_{3}$ and $\mathrm{YFeO}_{3}$: {A} comparative study},
    author = {Coutinho, P. V. and Cunha, F and Barrozo, Petrucio},
    journal = {Solid State Commun.},
    volume = {252},
    pages = {59--63},
    year = {2017},
    publisher = {Elsevier},
    doi = {10.1016/j.ssc.2017.01.019},
    url = {https://doi.org/10.1016/j.ssc.2017.01.019}
}

@book{born2013principles,
    title = {Principles of Optics: Electromagnetic Theory of Propagation, Interference and Diffraction of Light},
    author = {Born, Max and Wolf, Emil},
    year = {2013},
    publisher = {Elsevier, Amsterdam},
    isbn = {9332881537}
}

@article{gervais1974anharmonicity,
    title = {Anharmonicity in several-polar-mode crystals: adjusting phonon self-energy of {LO} and {TO} modes in $\mathrm{Al}_{2}\mathrm{O}_{3}$ and $\mathrm{TiO}_{2}$ to fit infrared reflectivity},
    author = {Gervais, Fran{\c{c}}ois and Piriou, Bernard},
    journal = {J. Phys. C: Solid State Phys.},
    volume = {7},
    number = {13},
    pages = {2374},
    year = {1974},
    publisher = {IOP Publishing},
    doi = {10.1088/0022-3719/7/13/017},
    url = {https://doi.org/10.1088/0022-3719/7/13/017}
}

@article{damen1966raman,
    title = {{R}aman {E}ffect in {Z}inc {O}xide},
    author = {Damen, T. C. and Porto, S. P. S. and Tell, B.},
    journal = {Phys. Rev.},
    volume = {142},
    issue = {2},
    pages = {570--574},
    numpages = {0},
    year = {1966},
    month = {Feb},
    publisher = {American Physical Society},
    doi = {10.1103/PhysRev.142.570},
    url = {https://link.aps.org/doi/10.1103/PhysRev.142.570}
}

@article{wojdyr2010fityk,
    title = {Fityk: a general-purpose peak fitting program},
    author = {Wojdyr, Marcin},
    journal = {	J. Appl. Crystallogr.},
    volume = {43},
    number = {5-1},
    pages = {1126--1128},
    year = {2010},
    publisher = {Wiley Online Library},
    doi = {10.1107/S0021889810030499},
    url = {https://doi.org/10.1107/S0021889810030499}
}

@article{gervais1983long,
    title = {Long-wavelength phonons in the four phases of $\{${$\mathrm{N}(\mathrm{CH}_{3})_{4}$}$\}_{2}\mathrm{CuCl}_{4}$ and effective charges},
    author = {Gervais, F. and Arend, H.},
    journal = {Z. Phys. B},
    volume = {50},
    number = {1},
    pages = {17--22},
    year = {1983},
    publisher = {Springer},
    doi = {10.1007/BF01307221},
    url = {https://doi.org/10.1007/BF01307221}
}

@article{kimel2004laser,
    title = {Laser-induced ultrafast spin reorientation in the antiferromagnet $\mathrm{TmFeO}_{3}$},
    author = {Kimel, A V and Kirilyuk, Andrei and Tsvetkov, Artem and Pisarev, R V and Rasing, {\relax Th}},
    journal = {Nature},
    volume = {429},
    number = {6994},
    pages = {850--853},
    year = {2004},
    publisher = {Nature Publishing Group},
    doi = {10.1038/nature02659},
    url = {https://doi.org/10.1038/nature02659}
}

@article{koshizuka1980inelastic,
    title = {Inelastic-light-scattering study of magnon softening in $\mathrm{ErFeO}_{3}$},
    author = {Koshizuka, N. and Ushioda, S.},
    journal = {Phys. Rev. B},
    volume = {22},
    issue = {11},
    pages = {5394--5399},
    numpages = {0},
    year = {1980},
    month = {Dec},
    publisher = {American Physical Society},
    doi = {10.1103/PhysRevB.22.5394},
    url = {https://link.aps.org/doi/10.1103/PhysRevB.22.5394}
}

@article{panchwanee2019temperature,
    title = {Temperature dependent dielectric and phonon study of polycrystalline $\mathrm{SmFeO}_{3}$},
    author = {Panchwanee, Anjali and Surampalli, Akash and Reddy, V Raghavendra},
    journal = {Phys. B: Condens. Matter.},
    volume = {570},
    pages = {187--190},
    year = {2019},
    publisher = {Elsevier},
    doi = {10.1016/j.physb.2019.06.035},
    url = {https://doi.org/10.1016/j.physb.2019.06.035}
}

@article{koshizuka1988raman,
    title = {{R}aman {S}cattering from {M}agnon {E}xcitations in $\mathrm{RFeO}_{3}$},
    author = {Koshizuka, Naoki and Hayashi, Kunihiko},
    journal = {J. Phys. Soc. Japan},
    volume = {57},
    number = {12},
    pages = {4418--4428},
    year = {1988},
    publisher = {The Physical Society of Japan},
    doi = {10.1143/JPSJ.57.4418},
    url = {https://doi.org/10.1143/JPSJ.57.4418}
}

@article{juraschek2017ultrafast,
    title = {{U}ltrafast {S}tructure {S}witching through {N}onlinear {P}hononics},
    author = {Juraschek, D. M. and Fechner, M. and Spaldin, N. A.},
    journal = {Phys. Rev. Lett.},
    volume = {118},
    issue = {5},
    pages = {054101},
    numpages = {5},
    year = {2017},
    month = {Jan},
    publisher = {American Physical Society},
    doi = {10.1103/PhysRevLett.118.054101},
    url = {https://link.aps.org/doi/10.1103/PhysRevLett.118.054101}
}

@article{loudon2001raman,
    title = {The {R}aman effect in crystals},
    author = {Loudon, R},
    journal = {Adv. Phys.},
    volume = {50},
    number = {7},
    pages = {813--864},
    year = {2001},
    publisher = {Taylor \& Francis},
    doi = {10.1080/00018730110101395},
    url = {https://doi.org/10.1080/00018730110101395}
}

@article{maschio2012infrared,
    title = {Ab initio analytical infrared intensities for periodic systems through a coupled perturbed {H}artree-{F}ock/{K}ohn-{S}ham method},
    author = {Maschio, Lorenzo and Kirtman, Bernard and Orlando, Roberto and R{\`e}rat, Michel},
    journal = {J. Chem. Phys.},
    volume = {137},
    number = {20},
    pages = {204113},
    year = {2012},
    publisher = {American Institute of Physics},
    doi = {10.1063/1.4767438},
    url = {https://doi.org/10.1063/1.4767438}
}

@article{maschio2013raman,
    title = {Ab initio analytical {R}aman intensities for periodic systems through a coupled perturbed {H}artree-{F}ock/{K}ohn-{S}ham method in an atomic orbital basis. {I}. {T}heory},
    author = {Maschio, Lorenzo and Kirtman, Bernard and R{\'e}rat, Michel and Orlando, Roberto and Dovesi, Roberto},
    journal = {J. Chem. Phys.},
    volume = {139},
    number = {16},
    pages = {164101},
    year = {2013},
    publisher = {American Institute of Physics},
    doi = {10.1063/1.4824442},
    url = {https://doi.org/10.1063/1.4824442}
}

@article{martin2001melting,
    title = {{M}elting of the cooperative {J}ahn-{T}eller distortion in $\mathrm{LaMnO}_{3}$ single crystal studied by {R}aman spectroscopy},
    author = {Mart{\'\i}n-Carr{\'o}n, L and de Andr{\'e}s, A},
    journal = {Eur. Phys. J. B},
    volume = {22},
    number = {1},
    pages = {11--16},
    year = {2001},
    publisher = {Springer},
    doi = {10.1007/PL00011129},
    url = {https://doi.org/10.1007/PL00011129}
}

@book{schubert2004infrared,
  title = {{I}nfrared {E}llipsometry on {S}emiconductor {L}ayer {S}tructures: {P}honons, {P}lasmons, and {P}olaritons},
  author = {Schubert, Mathias},
  volume = {209},
  year = {2004},
  publisher = {Springer, Berlin}
}

@article{bousquet2016non,
    title = {Non-collinear magnetism in multiferroic perovskites},
    author = {Bousquet, Eric and Cano, Andres},
    journal = {J. Phys.: Condens. Matter},
    volume = {28},
    pages = {123001},
    year = {2016},
    publisher = {IOP Publishing},
    doi = {10.1088/0953-8984/28/12/123001},
    url = {https://doi.org/10.1088/0953-8984/28/12/123001}
}

@inbook{balbashov1995submillimeter,
    title = {Submillimeter spectroscopy of antiferromagnetic dielectrics: Rare-earth orthoferrites, in High Frequency Processes in Magnetic Materials},
    editor = {Srinivasan, Gopalan and Slavin, Andrei N},
    author = {Balbashov, A M and Kozlov, G V and Mukhin, A A and Prokhorov, A S},
    journal = {High Frequency Processes in Magnetic Materials},
    volume = {56},
    year = {1995},
    publisher = {World Scientific Publishing, Singapore}
}

@article{stanislavchuk2016magnon,
    title = {Magnon and electromagnon excitations in multiferroic $\mathrm{DyFeO}_{3}$},
    author = {Stanislavchuk, T. N. and Wang, Yazhong and Janssen, Y. and Carr, G. L. and Cheong, S.-W. and Sirenko, A. A.},
    journal = {Phys. Rev. B},
    volume = {93},
    issue = {9},
    pages = {094403},
    numpages = {12},
    year = {2016},
    month = {Mar},
    publisher = {American Physical Society},
    doi = {10.1103/PhysRevB.93.094403},
    url = {https://link.aps.org/doi/10.1103/PhysRevB.93.094403}
}

@article{ivanov2023observation,
    title = {{O}bservation of {M}agnetic-{F}ield-{I}nduced {E}lectric {P}olarization in {T}erbium {O}rthoferrite},
    author = {Ivanov, V. {\relax Yu} and Kuzmenko, A. M. and Tikhanovskii, A. {\relax Yu} and Pronin, A. A. and Mukhin, A. A.},
    journal = {JETP Lett.},
    volume = {117},
    pages = {38–43},
    year = {2023},
    publisher = {Springer},
    doi = {10.1134/S0021364022602809},
    url = {https://doi.org/10.1134/S0021364022602809}
}

@article{dolg1989energy,
    title = {Energy-adjusted pseudopotentials for the rare earth elements},
    author = {Dolg, M and Stoll, H and Savin, A and Preuss, H},
    journal = {Theor. Chim. Acta},
    volume = {75},
    pages = {173--194},
    year = {1989},
    publisher = {Springer},
    doi = {10.1007/BF00528565},
    url = {https://doi.org/10.1007/BF00528565}
}

@article{dolg1993combination,
    title = {A combination of quasirelativistic pseudopotential and ligand field calculations for lanthanoid compounds},
    author = {Dolg, M and Stoll, H and Preuss, H},
    journal = {Theor. Chim. Acta},
    volume = {85},
    pages = {441--450},
    year = {1993},
    publisher = {Springer},
    doi = {10.1007/BF01112983},
    url = {https://doi.org/10.1007/BF01112983}
}

@article{yang2005valence,
    title = {Valence basis sets for lanthanide 4f-in-core pseudopotentials adapted for crystal orbital ab initio calculations},
    author = {Yang, Jun and Dolg, Michael},
    journal = {Theor. Chem. Acc.},
    volume = {113},
    pages = {212--224},
    year = {2005},
    publisher = {Springer},
    doi = {10.1007/s00214-005-0629-0},
    url = {https://doi.org/10.1007/s00214-005-0629-0}
}

@article{peintinger2013consistent,
    title = {Consistent Gaussian basis sets of triple-zeta valence with polarization quality for solid-state calculations},
    author = {Peintinger, Michael F and Oliveira, Daniel Vilela and Bredow, Thomas},
    journal = {J. Comp. Chem.},
    volume = {34},
    number = {6},
    pages = {451--459},
    year = {2013},
    publisher = {Wiley Online Library},
    doi = {10.1002/jcc.23153},
    url = {https://doi.org/10.1002/jcc.23153}
}

@article{smejkal2022emerging,
    title = {{E}merging {R}esearch {L}andscape of {A}ltermagnetism},
    author = {{\v{S}}mejkal, Libor and Sinova, Jairo and Jungwirth, Tomas},
    journal = {Phys. Rev. X},
    volume = {12},
    issue = {4},
    pages = {040501},
    numpages = {27},
    year = {2022},
    month = {Dec},
    publisher = {American Physical Society},
    doi = {10.1103/PhysRevX.12.040501},
    url = {https://link.aps.org/doi/10.1103/PhysRevX.12.040501}
}

@article{ivanov2023metamagnetic,
    title = {Metamagnetic and orientational transitions in $\mathrm{TbFeO}_{3}$ orthoferrite: magnetoelectric phase diagrams},
    author = {Ivanov, V. {\relax Yu} and Kuzmenko, A. M. and Tikhanovskii, A. {\relax Yu} and Mukhin, A A},
    journal = {Eur. Phys. J. Plus},
    volume = {138},
    number = {9},
    pages = {1--11},
    year = {2023},
    publisher = {Springer},
    doi = {10.1140/epjp/s13360-023-04422-2},
    url = {https://doi.org/10.1140/epjp/s13360-023-04422-2}
}

@article{komandin2023electric,
    title = {{E}lectric-dipole and magnetic absorption in $\mathrm{TbFeO}_{3}$ single crystals in the {THz}--{IR} range},
    author = {Komandin, Gennadiy A and Kuzmenko, Artem M and Spektor, Igor E and Mukhin, Alexander A},
    journal = {J. Appl. Phys.},
    volume = {133},
    pages = {194101},
    year = {2023},
    publisher = {AIP Publishing},
    doi = {10.1063/5.0149872},
    url = {https://doi.org/10.1063/5.0149872}
}

@article{zhang2024upconversion,
    title = {Terahertz-field-driven magnon upconversion in an antiferromagnet},
    author = {Zhang, Zhuquan and Gao, Frank Y and Chien, Yu-Che and Liu, Zi-Jie and Curtis, Jonathan B and Sung, Eric R and Ma, Xiaoxuan and Ren, Wei and Cao, Shixun and Narang, Prineha and von Hoegen, Alexander and Baldini, Edoardo and Nelson, Keith A.},
    journal = {Nature Phys.},
    volume = {20},
    pages = {788--793},
    year = {2024},
    publisher = {Nature Publishing Group UK London},
    doi = {10.1038/s41567-024-02386-3},
    url = {https://doi.org/10.1038/s41567-024-02386-3}
}

@article{zhang2024coupling,
    title = {Terahertz field-induced nonlinear coupling of two magnon modes in an antiferromagnet},
    author = {Zhang, Zhuquan and Gao, Frank Y and Curtis, Jonathan B and Liu, Zi-Jie and Chien, Yu-Che and von Hoegen, Alexander and Wong, Man Tou and Kurihara, Takayuki and Suemoto, Tohru and Narang, Prineha and Baldini, Edoardo and Nelson, Keith A.},
    journal = {Nature Phys.},
    volume = {20},
    pages = {801--806},
    year = {2024},
    publisher = {Nature Publishing Group UK London},
    doi = {10.1038/s41567-024-02386-3},
    url = {https://doi.org/10.1038/s41567-024-02386-3}
}

@article{huang2024extreme,
    title = {Extreme terahertz magnon multiplication induced by resonant magnetic pulse pairs},
    author = {Huang,  C. and Luo,  L. and Mootz,  M. and Shang,  J. and Man,  P. and Su,  L. and Perakis,  I. E. and Yao,  Y. X. and Wu,  A. and Wang,  J.},
    journal = {Nature Commun.},
    volume = {15},
    pages = {3214},
    year = {2024},
    publisher = {Nature Publishing Group UK London},
    doi = {10.1038/s41467-024-47471-6},
    url = {http://dx.doi.org/10.1038/s41467-024-47471-6},
}

@article{lyddane1941polar,
    title = {{O}n the {P}olar {V}ibrations of {A}lkali {H}alides},
    author = {Lyddane, R. H. and Sachs, R. G. and Teller, E.},
    journal = {Phys. Rev.},
    volume = {59},
    issue = {8},
    pages = {673--676},
    numpages = {0},
    year = {1941},
    month = {Apr},
    publisher = {American Physical Society},
    doi = {10.1103/PhysRev.59.673},
    url = {https://link.aps.org/doi/10.1103/PhysRev.59.673}
}

@article{leenders2024canted,
    title = {Canted spin order as a platform for ultrafast conversion of magnons},
    author = {Leenders, R. A. and Afanasiev, D. and Kimel, A. V. and Mikhaylovskiy, R. V.},
    journal = {Nature},
    volume = {630},
    pages = {335–339},
    year = {2024},
    publisher = {Nature Publishing Group},
    doi = {10.1038/s41586-024-07448-3},
    url = {https://doi.org/10.1038/s41586-024-07448-3}
}

@article{gomes2024lattice,
    title = {Lattice excitations in $\mathrm{NdFeO}_{3}$ through polarized optical spectroscopies},
    author = {Gomes, M. M. and Vilarinho, R. and Zhao, H. and Íñiguez-González, J. and Mihalik, M. and {Mihalik Jr.}, M. and Maia, A. and Goian, V. and Nuzhnyy, D. and Kamba, S. and {Agostinho Moreira}, J},
    journal = {Sci. Rep.},
    volume = {14},
    pages = {15378},
    year = {2024},
    publisher = {Springer Science and Business Media LLC},
    doi = {10.1038/s41598-024-66594-w},
    url = {http://dx.doi.org/10.1038/s41598-024-66594-w}
}

@article{dubrovin2024lattice,
    title = {Lattice dynamics and mixing of polar phonons in the rare-earth orthoferrite $\mathrm{TbFeO}_{3}$},
    author = {Dubrovin, R. M. and Roginskii, E. M. and Chernyshev, V. A. and Novikova, N. N. and Elistratova, M. A. and Eliseyev, I. A. and Smirnov, A. N. and Brulev, A. I. and Boldyrev, K. N. and Davydov, V. {\relax Yu.} and Mikhaylovskiy, R. V. and Kalashnikova, A. M. and Pisarev, R. V.},
    journal = {Phys. Rev. B},
    volume = {110},
    issue = {13},
    pages = {134310},
    numpages = {16},
    year = {2024},
    month = {Oct},
    publisher = {American Physical Society},
    doi = {10.1103/PhysRevB.110.134310},
    url = {https://link.aps.org/doi/10.1103/PhysRevB.110.134310}
}

@article{shapiro1974neutron,
    title = {Neutron-scattering studies of spin waves in rare-earth orthoferrites},
    author = {Shapiro, S. M. and Axe, J. D. and Remeika, J. P.},
    journal = {Phys. Rev. B},
    volume = {10},
    issue = {5},
    pages = {2014--2021},
    numpages = {0},
    year = {1974},
    month = {Sep},
    publisher = {American Physical Society},
    doi = {10.1103/PhysRevB.10.2014},
    url = {https://link.aps.org/doi/10.1103/PhysRevB.10.2014}
}

@article{venugopalan1983raman,
    title = {Raman scattering study of magnons at the spin-reorientation transitions of $\mathrm{TbFeO}_{3}$ and $\mathrm{TmFeO}_{3}$},
    author = {Venugopalan, S. and Dutta, Mitra and Ramdas, A. K. and Remeika, J. P.},
    journal = {Phys. Rev. B},
    volume = {27},
    issue = {5},
    pages = {3115--3118},
    numpages = {0},
    year = {1983},
    month = {Mar},
    publisher = {American Physical Society},
    doi = {10.1103/PhysRevB.27.3115},
    url = {https://link.aps.org/doi/10.1103/PhysRevB.27.3115}
}

@article{baryakhtar1983light,
    title = {Light scattering by magnons and magneto-optical effects in multisublattice magnets},
    author = {Bar'yakhtar, V G and Pashkevich, {\relax Yu} G and Sobolev, V L},
    journal = {Sov. Phys. JETP},
    volume = {58},
    pages = {945},
    year = {1983},
    url = {http://jetp.ras.ru/cgi-bin/dn/e_058_05_0945.pdf}
}

@article{cracknell1969scattering,
    title = {Scattering matrices for the {R}aman effect in magnetic crystals},
    author = {Cracknell, A P},
    journal = {J. Phys. C Solid State Phys.},
    volume = {2},
    number = {3},
    pages = {500},
    year = {1969},
    publisher = {IOP Publishing},
    doi = {10.1088/0022-3719/2/3/314},
    url = {https://doi.org/10.1088/0022-3719/2/3/314}
}

@article{wu2017crystal,
    title = {Crystal growth of $\mathrm{Sm}_{0.3}\mathrm{Tb}_{0.7}\mathrm{FeO}_{3}$ and spin reorientation transition in $\mathrm{Sm}_{1-x}\mathrm{Tb}_{x}\mathrm{FeO}_{3}$ orthoferrite},
    author = {Wu, Anhua and Wang, Bo and Zhao, Xiangyang and Xie, Tao and Man, Peiwen and Su, Liangbi and Kalashnikova, A M and Pisarev, R V},
    journal = {J. Magn. Magn. Mater.},
    volume = {426},
    pages = {721--724},
    year = {2017},
    publisher = {Elsevier},
    doi = {10.1016/j.jmmm.2016.10.136},
    url = {https://doi.org/10.1016/j.jmmm.2016.10.136}
}

@article{raut2018grain,
    title = {Grain boundary-dominated electrical conduction and anomalous optical-phonon behaviour near the Neel temperature in $\mathrm{YFeO}_{3}$ ceramics},
    author = {Raut, Subhajit and Babu, P D and Sharma, R K and Pattanayak, Ranjit and Panigrahi, Simanchalo},
    journal = {J. Appl. Phys.},
    volume = {123},
    number = {17},
    year = {2018},
    publisher = {AIP Publishing},
    doi = {10.1063/1.5012003},
    url = {https://doi.org/10.1063/1.5012003}
}

@article{nagrare2018hyperfine,
    title = {Hyperfine interaction, {R}aman and magnetic study of $\mathrm{YFeO}_{3}$ nanocrystals},
    author ={Nagrare, Bhagyashree S and Kekade, Shankar S and Thombare, Balu and Reddy, Raghavendra V and Patil, Shankar I},
    journal = {Solid State Commun.},
    volume ={280},
    pages = {32--38},
    year = {2018},
    publisher = {Elsevier},
    doi = {10.1016/j.ssc.2018.06.004},
    url = {https://doi.org/10.1016/j.ssc.2018.06.004}
}

@article{gupta2018temperature,
    title = {Temperature dependent Raman scattering and electronic transitions in rare earth $\mathrm{SmFeO}_{3}$},
    author = {Gupta, Surbhi and Medwal, Rohit and Pavunny, Shojan P and Sanchez, Dilsom and Katiyar, Ram S},
    journal = {Ceram. Int.},
    volume = {44},
    number = {4},
    pages = {4198--4203},
    year = {2018},
    publisher = {Elsevier},
    doi = {10.1016/j.ceramint.2017.11.223},
    url = {https://doi.org/10.1016/j.ceramint.2017.11.223}
}

@article{warshi2020cluster,
    title = {Cluster glass behavior in orthorhombic $\mathrm{SmFeO}_{3}$ perovskite: {I}nterplay between spin ordering and lattice dynamics},
    author = {Warshi, M Kamal and Kumar, Anil and Sati, Aanchal and Thota, S and Mukherjee, K and Sagdeo, Archna and Sagdeo, Pankaj R},
    journal = {Chem. Mater.},
    volume = {32},
    number = {3},
    pages = {1250--1260},
    year = {2020},
    publisher = {ACS Publications},
    doi = {10.1021/acs.chemmater.9b04703},
    url = {https://doi.org/10.1021/acs.chemmater.9b04703}
}

@article{turov1983broken,
    title = {Broken symmetry and magnetoacoustic effects in ferro and antiferromagnetics},
    author = {Turov, Evgenii A and Shavrov, Vladimir G},
    journal = {Sov. Phys. Usp.},
    volume = {26},
    number = {7},
    pages = {593},
    year = {1983},
    publisher = {IOP Publishing},
    doi = {10.1070/PU1983v026n07ABEH004449},
    url = {https://doi.org/10.1070/PU1983v026n07ABEH004449}
}

@article{zhang2024spin,
    title = {Spin switching in $\mathrm{Sm}_{0.7}\mathrm{Er}_{0.3}\mathrm{FeO}_{3}$ triggered by terahertz magnetic-field pulses},
    author = {Zhang, Zhenya and Kanega, Minoru and Maruyama, Kei and Kurihara, Takayuki and Nakajima, Makoto and Tachizaki, Takehiro and Sato, Masahiro and Kanemitsu, Yoshihiko and Hirori, Hideki},
    journal = {Nature Mater.},
    volume = {24},
    pages = {219--225},
    year = {2025},
    publisher = {Nature Publishing Group UK London},
    doi = {10.1038/s41563-024-02034-4},
    url = {https://doi.org/10.1038/s41563-024-02034-4}
}

@article{gareev2024optical,
    title = {{O}ptical {E}xcitation of {C}oherent {THz} {D}ynamics of the {R}are-{E}arth {L}attice through {R}esonant {P}umping of $f$--$f$ {E}lectronic {T}ransition in a {C}omplex {P}erovskite $\mathrm{DyFeO}_{3}$},
    author = {Gareev, T. T. and Sasani, A. and Khusyainov, D. I. and Bousquet, E. and Gareeva, Z. V. and Kimel, A. V. and Afanasiev, D.},
    journal = {Phys. Rev. Lett.},
    volume = {133},
    issue = {24},
    pages = {246901},
    numpages = {7},
    year = {2024},
    month = {Dec},
    publisher = {American Physical Society},
    doi = {10.1103/PhysRevLett.133.246901},
    url = {https://link.aps.org/doi/10.1103/PhysRevLett.133.246901}
}

@article{pavlovska2016thermal,
    title = {Thermal behaviour of $\mathrm{Sm}_{0.5}\mathrm{R}_{0.5}\mathrm{FeO}_{3}$ ($\mathrm{R}$ = $\mathrm{Pr}$, $\mathrm{Nd}$) probed by high-resolution {X}-ray synchrotron powder diffraction},
    author = {Pavlovska, Olena and Vasylechko, Leonid and Buryy, Oleh},
    journal = {Nanoscale Res. Lett.},
    volume = {11},
    pages = {1--6},
    year = {2016},
    publisher = {Springer},
    doi = {10.1186/s11671-016-1328-6},
    url = {https://doi.org/10.1186/s11671-016-1328-6}
}

@article{sasani2022origin,
    title = {Origin of nonlinear magnetoelectric response in rare-earth orthoferrite perovskite oxides},
    author = {Sasani, Alireza and \'I\~niguez, Jorge and Bousquet, Eric},
    journal = {Phys. Rev. B},
    volume = {105},
    issue = {6},
    pages = {064414},
    numpages = {8},
    year = {2022},
    month = {Feb},
    publisher = {American Physical Society},
    doi = {10.1103/PhysRevB.105.064414},
    url = {https://link.aps.org/doi/10.1103/PhysRevB.105.064414}
}

@article{pyatakov2012magnetoelectric,
    title = {Magnetoelectric and multiferroic media},
    author = {Pyatakov, Aleksandr P and Zvezdin, Anatolii K},
    journal = {Phys.-Usp.},
    volume = {55},
    number = {6},
    pages = {557},
    year = {2012},
    publisher = {IOP Publishing},
    doi = {10.3367/UFNe.0182.201206b.0593},
    url = {https://doi.org/10.3367/UFNe.0182.201206b.0593}
}

@article{fitzky2021ultrafast,
    title = {{U}ltrafast {C}ontrol of {M}agnetic {A}nisotropy by {R}esonant {E}xcitation of $4f$ {E}lectrons and {P}honons in $\mathrm{Sm}_{0.7}\mathrm{Er}_{0.3}\mathrm{FeO}_{3}$},
    author = {Fitzky, Gabriel and Nakajima, Makoto and Koike, Yohei and Leitenstorfer, Alfred and Kurihara, Takayuki},
    journal = {Phys. Rev. Lett.},
    volume = {127},
    issue = {10},
    pages = {107401},
    numpages = {6},
    year = {2021},
    month = {Sep},
    publisher = {American Physical Society},
    doi = {10.1103/PhysRevLett.127.107401},
    url = {https://link.aps.org/doi/10.1103/PhysRevLett.127.107401}
}

@article{wu2018growth,
    title = {Growth, and magnetic study of $\mathrm{Sm}_{0.4}\mathrm{Er}_{0.6}\mathrm{FeO}_{3}$ single crystal grown by optical floating zone technique},
    author = {Wu, Anhua and Zhao, Xiangyang and Man, Peiwen and Su, Liangbi and Kalashnikova, A M and Pisarev, R V},
    journal = {J. Cryst. Growth},
    volume = {486},
    pages = {169--172},
    year = {2018},
    publisher = {Elsevier},
    doi = {10.1016/j.jcrysgro.2018.01.026},
    url = {https://doi.org/10.1016/j.jcrysgro.2018.01.026}
}

@article{yamaguchi1974theory,
    title = {Theory of spin reorientation in rare-earth orthochromites and orthoferrites},
    author = {Yamaguchi, Tsuyoshi},
    journal = {J. Phys. Chem. Solids},
    volume = {35},
    number = {4},
    pages = {479--500},
    year = {1974},
    publisher = {Elsevier},
    doi = {10.1016/S0022-3697(74)80003-X},
    url = {https://doi.org/10.1016/S0022-3697(74)80003-X}
}

@article{li2025recent,
    author = {Qixin Li and Jie Zhou and Jiamin Shang and Hui Shen and Leifan Li and Fei Wang and Hao Wang and Xuanbing Shen and Tian Tian and A.M. Kalashinikova and Anhua Wu and Jiayue Xu},
    title = {Recent advances of rare earth orthoferrite $\mathrm{RFeO}_{3}$ magneto-optical single crystals},
    journal = {J. Cryst. Growth},
    year = {2025},
    volume = {649},
    publisher = {Elsevier},
    month = {jan},
    url = {https://linkinghub.elsevier.com/retrieve/pii/S0022024824003774},
    pages = {127939},
    doi = {10.1016/j.jcrysgro.2024.127939}
}

@article{vovk2025theory,
    title = {{T}heory of terahertz-driven magnetic switching in rare-earth orthoferrites: {T}he case of $\mathrm{TmFeO}_{3}$},
    author = {Vovk, N. R. and Ezerskaya, E. V. and Mikhaylovskiy, R. V.},
    journal = {Phys. Rev. B},
    volume = {111},
    issue = {6},
    pages = {064411},
    numpages = {17},
    year = {2025},
    month = {Feb},
    publisher = {American Physical Society},
    doi = {10.1103/PhysRevB.111.064411},
    url = {https://link.aps.org/doi/10.1103/PhysRevB.111.064411}
}

@article{kimel2024optical,
    title = {Optical read-out and control of antiferromagnetic N{\'e}el vector in altermagnets and beyond},
    author = {Kimel, A V and Rasing, {\relax Th} and Ivanov, B A},
    journal = {J. Magn. Magn. Mater.},
    volume = {598},
    pages = {172039},
    year = {2024},
    publisher = {Elsevier},
    doi = {10.1016/j.jmmm.2024.172039},
    url = {https://doi.org/10.1016/j.jmmm.2024.172039}
}

@article{naka2025altermagnetic,
    title = {Altermagnetic perovskites},
    author = {Naka, Makoto and Motome, Yukitoshi and Seo, Hitoshi},
    journal = {npj Spintronics},
    volume = {3},
    number = {1},
    pages = {1},
    year = {2025},
    publisher = {Nature Publishing Group UK London},
    doi = {10.1038/s44306-024-00066-9},
    url = {https://doi.org/10.1038/s44306-024-00066-9}
}

@article{song2025altermagnets,
    title = {Altermagnets as a new class of functional materials},
    author = {Song, Cheng and Bai, Hua and Zhou, Zhiyuan and Han, Lei and Reichlova, Helena and {Hugo Dil}, J  and Liu, Junwei and Chen, Xianzhe and Pan, Feng},
    journal = {Nature Rev. Mater.},
    pages = {1--13},
    year = {2025},
    publisher = {Nature Publishing Group UK London},
    doi = {10.1038/s41578-025-00779-1},
    url = {https://doi.org/10.1038/s41578-025-00779-1}
}

@article{baierl2016nonlinear,
    title = {Nonlinear spin control by terahertz-driven anisotropy fields},
    author = {Baierl, Sebastian and Hohenleutner, Matthias and Kampfrath, Tobias and Zvezdin, A K and Kimel, A V and Huber, Rupert and Mikhaylovskiy, R V},
    journal = {Nature Photon.},
    volume = {10},
    number = {11},
    pages = {715--718},
    year = {2016},
    publisher = {Nature Publishing Group UK London},
    doi = {10.1038/nphoton.2016.181},
    url = {https://doi.org/10.1038/nphoton.2016.181}
}

@article{zhao2016spin,
    title = {Spin reorientation transition in $\mathrm{Sm}_{0.5}\mathrm{Tb}_{0.5}\mathrm{FeO}_{3}$ orthoferrite single crystal},
    author = {Zhao, Xiangyang and Zhang, Kailin and Liu, Xiumei and Wang, Bo and Xu, Kai and Cao, Shixun and Wu, Anhua and Su, Liangbi and Ma, Guohong},
    journal = {AIP Adv.},
    volume = {6},
    number = {1},
    year = {2016},
    publisher = {AIP Publishing},
    doi = {10.1063/1.4939697},
    url = {https://doi.org/10.1063/1.4939697}
}

@article{nikolov1994mossbauer,
    title = {A {M}ossbauer study of temperature-driven spin-reorientation transitions in $\mathrm{TbFeO}_{3}$},
    author = {Nikolov, O and Hall, I and Barilo, S N and Guretskii, S A},
    journal = {J. Phys. Condens. Matter},
    volume = {6},
    number = {20},
    pages = {3793},
    year = {1994},
    publisher = {IOP Publishing},
    doi = {10.1088/0953-8984/6/20/019},
    url = {https://doi.org/10.1088/0953-8984/6/20/019}
}

@article{wang2025continuous,
    title = {Continuous two spin reorientation transitions and spin flips along the b-axis in $\mathrm{Er}_{0.6}\mathrm{Gd}_{0.4}\mathrm{FeO}_{3}$ single crystal},
    author = {Wang, Haixu and Li, Qixin and Su, Liangbi and Kou, Huamin and Kalashnikova, A M and Wu, Anhua},
    journal = {Phys. B Condens. Matter},
    volume = {705},
    pages = {417109},
    year = {2025},
    publisher = {Elsevier},
    doi = {10.1016/j.physb.2025.417109},
    url = {https://doi.org/10.1016/j.physb.2025.417109}
}

@article{khokhlov2024double,
    title = {Double pulse all-optical coherent control of ultrafast spin-reorientation in an antiferromagnetic rare-earth orthoferrite},
    author = {Khokhlov, N E and Dolgikh, A E and Ivanov, B A and Kimel, A V},
    journal = {APL Mater.},
    volume = {12},
    number = {5},
    year = {2024},
    publisher = {AIP Publishing},
    doi = {10.1063/5.0197976},
    url = {https://doi.org/10.1063/5.0197976}
}

@article{li2019spin,
    title = {Spin switching in single crystal $\mathrm{PrFeO}_{3}$ and spin configuration diagram of rare earth orthoferrites},
    author = {Li, Enyu and Feng, Zhenjie and Kang, Baojuan and Zhang, Jincang and  Ren, Wei and Cao, Shixun},
    journal = {Journal of Alloys and Compounds},
    volume = {811},
    pages = {152043},
    year = {2019},
    publisher = {Elsevier},
    doi = {10.1016/j.jallcom.2019.152043},
    url = {https://doi.org/10.1016/j.jallcom.2019.152043}
}

@book{zvezdin1985rare,
    title = {Rare earth ions in magnetically ordered crystals},
    author = {Zvezdin, A K and Matveev, V M and Mukhin, A A and Popov, A I},
    journal = {Moscow Izdatel Nauka},
     publisher = {Nauka, Moscow},
    year = {1985}
}

@article{buchelnikov1996magnetoacoustics,
    title = {Magnetoacoustics of rare-earth orthoferrites},
    author = {Buchel'nikov, Vasilii D and Dan'shin, Nikolay Kuzmich and Tsymbal, L T and Shavrov, Vladimir G},
    journal = {Phys. Usp.},
    volume = {39},
    number = {6},
    pages = {547},
    year = {1996},
    publisher = {IOP Publishing},
    doi = {10.1070/PU1996v039n06ABEH000148},
    url = {https://doi.org/10.1070/PU1996v039n06ABEH000148}
}

@article{balbashov1989anomalies,
    title = {{A}nomalies of high-frequency magnetic properties and new orientational transitions in $\mathrm{HoFeO}_{3}$},
    author = {Balbashov, A M and Kozlov, G V and Lebedev, S P and Mukhin, A A and Pronin, A {\relax Yu} and Prokhorov, A S},
    journal = {Sov. Phys. JETP},
    volume = {68},
    number = {3},
    pages = {629--638},
    year = {1989},
    url = {http://www.jetp.ras.ru/cgi-bin/dn/e_068_03_0629.pdf}
}

@dataset{dataset,
    author = {Dubrovin, R M and Brulev, A I and Vovk, N R and Eliseyev, I A and Novikova, N N and Smirnov, A N and  Chernyshev, V A and Davydov, V {\relax Yu} and Wu, Anhua and Su, Liangbi and Mikhaylovskiy, R V and Kalashnikova, A M and Pisarev, R V},
    title = {Spin and lattice dynamics at the spin-reorientation transitions in the rare-earth orthoferrite $\mathrm{Sm}_{0.55}\mathrm{Tb}_{0.45}\mathrm{FeO}_{3}$ [{D}ata set]},
    month = jul,
    year = 2025,
    publisher = {Zenodo},
    doi = {10.5281/zenodo.17342009},
    url = {https://doi.org/10.5281/zenodo.17342009},
}
\end{document}